\documentclass[11pt]{article}
\usepackage{jheppubmod}
\usepackage{color}
\usepackage[showonlyrefs]{mathtools}
\usepackage{psfrag}
\usepackage{array}
\usepackage{amssymb}
\usepackage{graphicx}
\usepackage{amsmath}
\usepackage{empheq}
\usepackage{slashed}
\usepackage{caption}
\usepackage[labelsep=quad]{subcaption}
\usepackage{epstopdf}
	
\usepackage{epsfig}
\usepackage[punctsep]{collref}
\newcommand{\be}{\begin{equation}}
\newcommand{\ee}{\end{equation}}
\collectsep[]{;}	
\usepackage{cite}
  
\allowdisplaybreaks

\def\AA{{\cal A}}

\def\CC{{\cal C}}
\def\DD{{\cal D}}

\def\KK{{\cal K}}

\def\MM{{\cal M}}
\def\NN{{\cal N}}
\def\OO{{\cal O}}

\def\SS{{\cal S}}

\def\VV{{\cal V}}

\def\tts{{$tt^*$ }}

\def\Tr{{\rm {Tr}}}

\def\gym{g_{YM}}
\def\barM{\overline}
\def\beq{\begin{equation}}
\def\eeq{\end{equation}}
\newcommand{\bea}{\begin{eqnarray}}
\newcommand{\eea}{\end{eqnarray}}

\title{
On exact correlation functions in $SU(N)$ $\NN=2$ superconformal QCD}

\author[a]{Marco Baggio,}
\author[b]{Vasilis Niarchos,}
\author[c,d]{Kyriakos Papadodimas}
\affiliation[a]{Institut fur Theoretische Physik, ETH Zurich,
CH-8093 Zurich, Switzerland.}

\affiliation[b]{Crete Center for Theoretical Physics
and Crete Center for Quantum Complexity and Nanotechnology,
Department of Physics, University of Crete, 71303, Greece.}
\affiliation[c]{Theory Group, Physics Department, CERN, CH-1211 Geneva 23, Switzerland.}
\affiliation[d]{Van Swinderen Institute for Particle Physics and Gravity, University of Groningen, Nijenborgh 4, 9747 AG Groningen, The Netherlands.}
\emailAdd{baggiom@ethz.ch}
\emailAdd{niarchos@physics.uoc.gr}
\emailAdd{kyriakos.papadodimas@cern.ch}

\preprint{CCQCN-2015-xx
\\ \hspace*{\fill}
CCTP-2015-xx\\\hspace*{\fill}
CERN-PH-TH-2015-190}

\date{}

\abstract{
We consider the exact coupling constant dependence of extremal correlation functions of $\NN=2$ chiral primary operators
in 4d $\NN=2$ superconformal gauge theories with gauge group $SU(N)$ and $N_f=2N$ massless fundamental hypermultiplets.
The 2- and 3-point functions, viewed as functions of the exactly marginal coupling constant and theta angle, obey
the \tts equations. In the case at hand, the \tts equations form a set of complicated non-linear
coupled matrix equations. We point out that there is an ad hoc self-consistent ansatz that reduces this set 
of partial differential equations to a sequence of decoupled semi-infinite Toda chains, similar to the one encountered 
previously in the special case of $SU(2)$ gauge group. This ansatz requires a surprising new non-renormalization 
theorem in $\NN=2$ superconformal field theories. We derive a general 3-loop perturbative formula for 2- and 3-point 
functions in the $\NN=2$ chiral ring of the $SU(N)$ theory, and in all explicitly computed examples we find agreement with 
the \tts equations, as well as the above-mentioned ansatz. This is suggestive evidence for an interesting non-perturbative 
conjecture about the structure of the $\NN=2$ chiral ring in this class of theories. We discuss several implications of this 
conjecture. For example, it implies that the holonomy of the vector bundles of chiral primaries over the superconformal 
manifold is reducible. It also implies that a specific subset of extremal correlation functions can be computed in the $SU(N)$ 
theory using information solely from the $S^4$ partition function of the theory obtained by supersymmetric localization. 
}


\begin{document}
\maketitle

\section{Introduction}
\label{intro}

The \tts equations provide a powerful set of relations between 2- and 3-point functions in the chiral ring of
$\NN=2$ theories. They were originally derived with the method of the topological-antitopological fusion 
in 2d $\NN=(2,2)$ theories in \cite{Cecotti:1991me}. In 4d $\NN=2$ superconformal field theories (SCFTs)
they were derived using superconformal Ward identities in conformal perturbation theory in 
\cite{Papadodimas:2009eu}.

There are important differences between $\NN=2$ chiral rings in two and four dimensions that are reflected 
 in the geometry of the superconformal manifold, as well as the structure and solutions of the \tts equations.
For example, 2d $\NN=(2,2)$ chiral rings have a spectrum with an upper bound on the scaling dimension
\cite{Lerche:1989uy}. On the other hand, it is believed that the $\NN=2$ chiral ring of a generic 
4d $\NN=2$ SCFT is freely generated without any upper bound on scaling dimensions. As we pointed out
in \cite{Baggio:2014ioa} this feature has important implications for the structure of the \tts equations. 

As a more explicit illustration of this fact, in specific 2d theories \cite{Cecotti:1991me,Cecotti:1991vb} the \tts 
equations take the form of a {\it periodic} Toda chain. In four-dimensional examples, e.g.\ the $SU(2)$ 
$\NN=2$ super-Yang-Mills (SYM) theory coupled to 4 hypermultiplets, also known as $SU(2)$ $\NN=2$ superconformal QCD (SCQCD),
which was analyzed in \cite{Baggio:2014ioa,Baggio:2014sna}, the \tts equations also reduce to a Toda chain, but in this case the chain 
is {\it semi-infinite}. Periodic and semi-infinite Toda chains have qualitatively different features. The solution of the periodic case is uniquely 
fixed by unitarity and a small set of perturbative data, but the solution of the semi-infinite case requires complete knowledge 
of a single member of the chain. The latter does not appear 
to be uniquely determined by consistency and a few perturbative data. In the context of the $SU(2)$ $\NN=2$ 
SCQCD theory we proposed \cite{Baggio:2014ioa,Baggio:2014sna} that the solution can be
determined from the exact form of the Zamolodchikov metric on the superconformal manifold,
which is the lowest non-trivial member of the Toda chain. In turn, the Zamolodchikov metric, and the exact quantum 
K\"ahler potential, are directly related to the $S^4$ partition function of the theory \cite{Gerchkovitz:2014gta,Gomis:2014woa},
which can be computed efficiently using localization techniques \cite{Pestun:2007rz}.

The clean example of the $SU(2)$ $\NN=2$ SCQCD theory invites us to think more generally about the structure
of the \tts equations in four-dimensional theories, the constraints that they impose on the correlation functions of 
the $\NN=2$ chiral ring, and the independent data needed to determine a physically consistent solution. 
The precise answer to many of these questions is far from obvious. For instance, already in the general 
$SU(N)$ SCQCD theory the \tts equations (in an appropriate gauge) take the form of an infinite set of coupled, non-linear 
differential equations for matrix-valued quantities
whose size grows indefinitely with the scaling dimension (see equation \eqref{introa4} below). 
In more general $\NN=2$ SCFTs, which possess higher dimensional superconformal manifolds, 
the structure of the \tts equations is an even more complicated set of partial differential equations.

As a step towards a better understanding of this structure, in this paper we initiate a more detailed study of the \tts equations of the general
$SU(N)$ $\NN=2$ SCQCD theory. First, by an explicit computation of chiral primary 2- and 3-point functions up to 3-loops in perturbation theory, we  
verify that the matrix-valued \tts equations \eqref{introa4} are satisfied up to that order. Second, 
we investigate a specific non-perturbative ansatz for the complete solution of these equations, which is consistent with 
the perturbative computations. The precise form of this ansatz will be explained in the next subsection.
One of its characteristic properties is that it leads to a drastic reduction of the complicated set of matrix-valued equations \eqref{introa4} 
to a decoupled set of semi-infinite Toda chains (similar to the chains encountered in the $SU(2)$ case). 
These can be solved recursively from a single member in each chain. 

Besides this drastic reduction of the \tts equations the proposed solution has other surprising properties. One of them is the requirement
of a novel non-renormalization theorem in this class of $\NN=2$ theories, where orthogonal chiral primary operators in a specific basis do not mix 
by quantum finite coupling effects. Relatedly, the holonomy of the vector bundles of chiral primaries over the superconformal 
manifold is required to be reducible. 

At the moment, we do not have a proof of the above-mentioned ansatz in gauge theory. Besides the favorable evidence 
provided by explicit 3-loop computations in perturbation theory, it is encouraging that this ansatz is mathematically a self-consistent
way to solve the exact non-perturbative \tts equations. Nevertheless, we do not have an argument that this is the only way to solve
the \tts equations and the logical possibility of more complicated alternatives (that we have not yet discovered) remains. We point out 
some alternatives in the main text. A complete non-perturbative proof of the no-mixing conjecture, the explanation of its physical
origin, and its relevance in more general $\NN=2$ theories are some of the interesting open questions that this work is opening up.

\subsection{Summary of main results}

The $\NN=2$ chiral primary fields of the $SU(N)$ $\NN=2$ SYM theory coupled to $2N$ hypermultiplets
are believed to be freely generated from the product of a finite set of $N-1$ generators. In the standard Lagrangian description of the theory
these generators are represented as single-trace operators $\Tr [\varphi^{\ell+1}]$, $\ell=1,\ldots,N-1$, where $\varphi$ is the adjoint complex scalar field 
in the $\NN=2$ vector multiplet. From now on, we will denote the generic chiral primary in this representation as
$\phi_K$ with a multi-index $K=\{ n_\ell \}$ 
\beq
\label{introa5}
\phi_{\{n_\ell \}} \propto \prod_{\ell=1}^{N-1} \left( \Tr [\varphi^{\ell+1} ] \right)^{n_\ell}
~.
\eeq
The anti-chiral primaries are multi-trace operators of the complex-conjugate field $\overline\varphi$ and will be denoted as 
$\overline\phi_{K}$. We single out the special chiral primary $\phi_2 \propto \Tr[\varphi^2]$, which is the single operator
with scaling dimension 2 in this family. The supersymmetric descendant $\OO_\tau = Q^4\cdot  \phi_2$ of this operator gives the 
exactly marginal interaction of the theory associated to the complexified coupling constant
$\tau = \frac{\theta}{2\pi} + i \frac{4\pi }{\gym^2}$. As usual, $\theta$ is the theta-angle of the theory and $\gym$ the 
gauge coupling. 

In a specific set of normalization conventions, the so-called holomorphic gauge \cite{Baggio:2014ioa},
the OPE coefficients $C_{KL}^M$ are 0 or 1, so that
\beq
\label{introa3}
\phi_K (x) \, \phi_L(0) = \phi_{K+L}(0)+\ldots
\eeq
Here $\phi_{K+L}$ denotes the multi-trace operator $:\phi_K\phi_L:$ and the dots represent descendants with higher scaling dimension. 
In these normalization conventions the \tts equations become a coupled set of matrix partial differential equations 
\beq
\label{introa4}
\partial_{\barM \tau} \left( g^{\barM M_\Delta L_\Delta} \partial_\tau g_{K_\Delta \barM M_\Delta} \right)
= g_{K_{\Delta+2},\barM R_\Delta+\barM 2}\, g^{\barM R_\Delta L_\Delta} 
- g_{K_\Delta \barM R_\Delta} \, g^{\barM R_\Delta-\barM 2, L_\Delta-2} - g_2 \, \delta_{K_\Delta}^{L_\Delta}
\eeq
for the 2-point function coefficients
\beq
\label{introaaa}
\langle \phi_K (x) \overline{\phi_L}(0) \rangle = \frac{g_{K\barM L}}{|x|^{2\Delta}}
~.
\eeq
Here $\Delta$ is the common scaling dimension of the insertions. The index 2 in the notation employed in \eqref{introa4} refers to the chiral primary 
$\phi_2$. (Further explanations of \eqref{introa4} are provided in section \ref{review}.)

In this paper we evaluate the relevant Feynman diagrams and derive a (3-loop) formula that computes $\OO(\gym^4)$ 
corrections to the general 2-point function $g_{K\overline{L}}$ in the $SU(N)$ $\NN=2$ chiral ring. We apply this formula 
in several explicit $SU(3)$ and $SU(4)$ examples and verify that the equations \eqref{introa4} are indeed
obeyed up to that order. 

Moreover, the perturbative results provide highly suggestive evidence that the chiral primary correlators in this class of theories may 
correspond to a solution of \eqref{introa4} of rather special form. The main aim of this paper is to describe the ansatz for this special 
solution and explore its implications for chiral primary correlators.   

Our discussion begins with the special role played by the chiral primary 
$\phi_2 \propto \Tr[\varphi^2]$. The \tts equations \eqref{introa4} relate the 2-point functions of chiral primaries at a given level of R-charge
with those chiral primaries that can be reached by the action of the special chiral primary $\phi_2$ or its conjugate $\overline{\phi}_2$. Hence it is
natural to consider the operator $C_2$ (and its conjugate $C_2^\dagger$), corresponding to chiral ring OPE multiplication by $\phi_2$, 
acting on the vector space of chiral primaries. We first show \emph{at tree level} that 
it is possible to construct a basis of chiral primaries that diagonalizes simultaneously the 2-point functions 
$g_{K\overline{L}}$ and the action of $C_2$.\footnote{By this we mean that $C_2$ sends an element of the basis to a single other element of the basis.} 
Rather surprisingly, we find that even after including 3-loop corrections to these correlators, 
there still exists a basis where the simultaneous diagonalization of 2-point functions and of $C_2$ is possible.

This encourages us to investigate the possibility that there is a basis of chiral primaries in which  
the full non-perturbative matrix of 2-point functions remains diagonal simultaneously with the matrix $C_2$ for all values of the coupling. 
This possibility lies at the core of the ansatz that we explore in this paper. 
The assumption that there is a basis in the chiral ring in which different degenerate operators 
do not mix under conformal perturbation theory has a simple geometric meaning. It is the geometric statement that the gauge connection on the 
chiral primary vector bundles over the superconformal manifold is reducible. More specifically, at a generic scaling 
dimension $\Delta$ with degeneracy $D$ the holonomy of the chiral primary vector bundle is, according to this ansatz, not $U(D)$ 
(as one might have a priori expected), but much smaller, $U(1)^D$. 

As we explain in the main text, the next-to-leading order perturbative results in this paper do not allow us to check conclusively
the no-mixing properties for all possible 2-point functions. They only allow us to find direct non-trivial
evidence of the absence of mixing for degenerate operators that contain `a different number of $\Tr\left[\varphi^2\right]$ factors'.
This leaves open the possibility of a partial mixing in gauge theory, where at generic scaling dimensions
the 2-point functions are non-perturbatively block-diagonal instead of completely diagonal. In that case, 
the chiral primary vector bundles over the superconformal manifold would be partially reducible to a product of $U(1)$ line bundles 
times bundles with a non-abelian connection. We explain when non-abelian factors in the holonomy could in principle appear.

Under the postulate of full reducibility the proposed solution implies that the \tts equations reduce to a \emph{decoupled} set of semi-infinite 
Toda chains. Each of these Toda chains, whose explicit form can be found in equation \eqref{eq:gentts}, can be solved in terms of a single external datum. 
A notable class of data that we can compute in this way are the 2-point functions of the form 
$\langle (\Tr[\varphi^2])^n (x) (\Tr[\overline{\varphi}^2])^n(0) \rangle$. Similar to the $SU(2)$ results in \cite{Baggio:2014ioa,Baggio:2014sna}, 
we find specific predictions for these data in terms of the Zamolodchikov metric of the $SU(N)$ theory, 
which is known exactly from supersymmetric localization \cite{Pestun:2007rz,Gomis:2014woa}. We further show that the no-mixing conjecture
allows to extract more information from the $S^4$ partition function for additional extremal correlation functions.

\subsection{Outline of the paper}

In section \ref{review} we review the basic features of the theory of interest and set up our notation. 
The precise form of the \tts equations that we analyze is also reviewed here. 
Section \ref{decoupling} explains the main proposal and how it leads to a recursive solution of the \tts 
equations in the $SU(N)$ theory. Starting at tree level we present a linear transformation on the 
vector space of chiral primaries that diagonalizes simultaneously the 2-point functions and the components
of the OPE coefficients $C_{2K}^L$, and reorganizes the \tts equations into a set of decoupled semi-infinite Toda chains.
We propose that the nice properties of this basis continue to hold non-perturbatively at finite coupling.
The implications of this proposal are discussed further 
in section \ref{implications}, which contains a list of specific predictions for 
exact correlation functions in chiral ring of the $SU(N)$ $\NN=2$ SCQCD theory. These predictions 
are tested non-trivially in perturbation theory in section \ref{perturbation} using a general $\OO(\gym^4)$
perturbative formula for 2-point functions in the chiral ring derived in appendix \ref{pert}. We conclude
in section \ref{outlook} with a summary of interesting open issues. 
For the benefit of the reader appendix \ref{explicit} contains a supplementary description of the 
diagonalization of 2-point functions discussed in section \ref{decoupling}.

\section{Review of the \tts equations in $SU(N)$ $\NN=2$ SCQCD theory}
\label{review}

The general properties of the \tts equations in four-dimensional $\NN=2$ SCFTs are reviewed in 
Ref.\ \cite{Baggio:2014ioa}, whose notation we will mostly follow. In the rest of the paper we will
omit many of the technical details, which can be found in \cite{Baggio:2014ioa}, and will focus directly on 
the case of interest: the $\NN=2$ superconformal QCD  (SCQCD) theory defined as $\NN=2$ super-Yang-Mills
theory with gauge group $SU(N)$ coupled to $2N$ massless hypermultiplets in the fundamental representation. The global symmetry group of this theory
for generic $N$ is $U(2N) \times SU(2)_R \times U(1)_R$, where $U(2N)$ is the flavor symmetry group and 
$SU(2)_R\times U(1)_R$ is the R-symmetry group.

The $\NN=2$ chiral ring of the $SU(N)$ $\NN=2$ SCQCD theory is freely generated by the $N-1$ single-trace
operators
\beq
\label{reviewaa}
\phi_{\ell+1}  \propto \Tr \left[ \varphi^{\ell+1} \right]~, ~~ \ell=1,2,\ldots, N-1
\eeq
with scaling dimension $\Delta=\ell+1$. Here, $\varphi$ is the adjoint complex scalar field in the $\NN=2$ vector
multiplet. Hence, the generic chiral primary field 
\beq
\label{reviewaaa}
\phi_{\{n_\ell \}} \propto \prod_{\ell=1}^{N-1} \left( \Tr [\varphi^{\ell+1} ] \right)^{n_\ell}
\eeq
is a multi-trace product of arbitrary powers of these generators. These fields are neutral under the flavor $U(2N)$
and the R-symmetry group $SU(2)_R$. They are charged under the $U(1)_R$ with R-charge $R$ and scaling
dimension
\beq
\label{reviewaab}
\Delta_{\{ n_\ell \}} = \frac{R_{\{ n_\ell \}}}{2} = \sum_{\ell = 1}^{N-1} (\ell+1)n_\ell
~.
\eeq
The $\NN=2$ SCQCD theory has a single (complex) exactly marginal deformation
\beq
\label{reviewab}
\delta S= \frac{\delta \tau}{4\pi^2} \int d^4x \, \OO_\tau(x) 
+ \frac{\delta \barM \tau}{4\pi^2} \int d^4x \, \overline{\OO}_\tau(x)
~,
\eeq
where $\OO_\tau$ is the supersymmetry descendant 
\beq
\label{reviewac}
\OO_\tau = Q^4 \cdot \phi_2
\eeq
of the chiral primary $\phi_2 \propto \Tr [\varphi^2]$. The notation $Q^4\cdot \phi_2$ denotes the nested
(anti)-commutator of four supercharges of left chirality on the field $\phi_2$.
The corresponding exactly marginal coupling that parametrizes the (complex) 1-dimensional superconformal 
manifold is the complexified gauge coupling constant $\tau = \frac{\theta}{2\pi} + i \frac{4\pi }{\gym^2}$.

In what follows we will employ a specific set of normalization conventions for the chiral primaries 
$\phi_{\{ n_\ell \}}$, where 
$(i)$ $\phi_2$ adheres to the conventions \eqref{reviewab}-\eqref{reviewac},
$(ii)$ we require
\beq
\label{reviewad}
\langle \OO_\tau (x) \overline \OO_\tau (0) \rangle = \nabla_x^2 \nabla^2_x 
\langle \phi_2(x) \overline\phi_2(0) \rangle
~,
\eeq
$(iii)$ the remaining generators in \eqref{reviewaa} are chosen with an arbitrary holomorphic normalization factor,
and $(iv)$ we require that the non-vanishing OPE coefficients are
\beq
\label{reviewae}
C_{K\, L}^{K+L}=1
~.
\eeq
For general indices $K, L,\ldots$ of the form $\{ k_\ell \} , \{l_\ell \},\ldots$ the notation $K+L$ in \eqref{reviewae} 
denotes the index $\{k_\ell+l_\ell \}$ and the OPE coefficient \eqref{reviewae} implies the operator product expansion
\beq
\label{reviewaf}
\phi_{\{k_\ell\}}(x) \, \phi_{\{ l_\ell \}}(0) = \phi_{\{ k_\ell+l_\ell \}}(0) +\cdots
~.
\eeq
This OPE is enough to fix the normalization of all multi-trace chiral primaries in terms of the normalization of the 
generators \eqref{reviewaa}.

As explained in \cite{Papadodimas:2009eu, Baggio:2014ioa} it is most appropriate to think of the chiral primary fields $\phi_L$ as sections in a 
holomorphic vector bundle $\VV$ whose base space is the superconformal manifold of the theory. 
The above set of conventions is a choice that makes the rescaled chiral primary fields
\beq
 e^{-\frac{R_L}{c'} \,\KK} \phi_L
\eeq
holomorphic sections of the bundle $\VV$. Here, $R_L$ is the $U(1)_R$ charge of the fields $\phi_L$,
$c' = 8 \times 192 \times c$ (where $c$ is the central charge of the CFT), and $\KK$ is the exact K\"ahler potential of
the superconformal manifold. $\phi_L$ are the chiral primaries in the conventions 
\eqref{reviewad}-\eqref{reviewae}. The reason for the appearance of the factor $e^{-\frac{R_L}{c'} \KK}$ can be traced
back to the choice of normalization conventions for the supercharges, or equivalently to the choice of a section 
in the holomorphic line bundle associated to the supercurrents.

These choices constitute the so-called holomorphic gauge where a connection $A$ on $\VV$ 
compatible with the 2-point function coefficients $g_{K\barM L}$ has components \cite{Papadodimas:2009eu}
\bea
\label{reviewaia}
\left( A_\tau \right)_K^{L} &=& g^{\barM M L}\partial_\tau g_{K\barM M} - \frac{2R_K}{c'} \partial_\tau \KK \, \delta_K^{L}
~,\\
\label{reviewaib}
\left( \barM A_{\barM \tau} \right)_{\barM K}^{\barM L} &=& g^{\barM L M}\partial_{\barM \tau} g_{M\barM K} 
- \frac{2R_{\barM K}}{c'} \partial_{\barM \tau} \KK \, \delta_{\barM K}^{\barM L}
~.
\eea
We remind the reader that $g_{K \barM L}$ is defined in \eqref{introaaa}. The notation $g^{\barM K L}$ refers to the components 
of the inverse matrix of 2-point function coefficients:
$g_{K\barM M} g^{\barM M L} = \delta_K^L$.
The components $(A_\tau)^{\barM K}_{\barM L}$, $(\barM A_{\barM \tau})^{K}_{L}$ vanish by definition in the holomorphic 
gauge.

The \tts equations express the curvature of this connection. In holomorphic gauge they lead to a 
set of partial differential equations for the 2- and 3-point function coefficients $g_{K\barM L}, C_{KL}^M$, which have the form  
 \bea
\label{reviewaj}
&&\partial_{\barM \tau} \left( g^{\barM M_\Delta L_\Delta} \partial_\tau g_{K_\Delta \barM M_\Delta} \right)
\\
&&= C_{2K_\Delta}^{P_{\Delta+2}} g_{P_{\Delta+2} \barM Q_{\Delta+2}}\, C_{\barM 2 \barM R_\Delta}^{*\barM Q_{\Delta+2}}   g^{\barM R_\Delta L_\Delta} 
- g_{K_\Delta \barM N_\Delta} \, C_{\barM 2 \barM U_{\Delta-2}}^{*\barM N_\Delta}  g^{\barM U_{\Delta- 2}  V_{\Delta-2}} C_{2 V_{\Delta-2}}^{L_\Delta}
- g_2 \, \delta_{K_\Delta}^{L_\Delta}
~,\nonumber
\eea
where $C_{2K}^L$ denotes the coefficient in the OPE of the chiral primaries $\phi_2$ and $\phi_K$.
In the conventions \eqref{reviewae} we set $C_{2K}^L = \delta^{L}_{K+2}$ and \eqref{reviewaj} simplifies to \eqref{introa4}.  
Note that in \eqref{reviewaj} we are using the more explicit index notation $K_\Delta, \ldots$ to keep track of the scaling dimension $\Delta$ of 
the corresponding chiral primary fields. In this way it is apparent that equation \eqref{reviewaj} is an equation that
relates 2-point functions at three different scaling dimensions: $\Delta-2$, $\Delta$, and $\Delta+2$. 

Finally, $g_2 = \langle \Tr[\varphi^2](1) \Tr[\overline{\varphi}^2](0)\rangle$ is related by \eqref{reviewad} to the Zamolodchikov
metric of the theory up to an overall constant factor. Hence $g_2$ is conveniently related \cite{Gerchkovitz:2014gta} to the $S^4$ 
partition function of the theory, $Z_{S^4}$, which is 
exactly computable as an $(N-1)$-dimensional ordinary integral with the use of supersymmetric localization 
methods \cite{Pestun:2007rz}. The precise relation between $g_2$ and $Z_{S^4}$ in our conventions is
\beq
\label{reviewak}
g_2 = \partial_\tau \partial_{\barM \tau} \log Z_{S^4}
~.
\eeq
This equation was first proven in \cite{Gerchkovitz:2014gta}.

In the special case of the $SU(2)$ theory, which was the focus of Ref.\ \cite{Baggio:2014sna}, the chiral ring
is freely generated by $\Tr[\varphi^2]$ only, and then \eqref{reviewae}, \eqref{reviewaj} reduce to the simple recursive set of differential equations 
\beq
\label{reviewaka}
\partial_{\barM \tau} \partial_\tau \log g_{2n} = \frac{g_{2n+2}}{g_{2n}} - \frac{g_{2n}}{g_{2n-2}} - g_2 ~, ~~n=1,\ldots
\eeq
where by definition $g_0=1$. These equations can be recast as a semi-infinite Toda chain
\beq
\label{reviewal}
\partial_{\barM \tau} \partial_{\tau} q_n = e^{q_{n+1}-q_n} - e^{q_n - q_{n-1}} ~, ~~n=1,\ldots
\eeq
by setting 
\beq
\label{reviewala}
g_{2n} 
\equiv \langle \left( \Tr[\varphi^2]\right)^n (1) \left(\Tr[\overline{\varphi}^2]\right)^n (0)\rangle
= \exp \left( q_n -\log Z_{S^4} \right)
~.
\eeq

In the general $SU(N)$ case, due to the presence of additional chiral ring generators \eqref{reviewaa}, the equations \eqref{reviewaj} 
are instead a complicated set of coupled differential equations for matrix-valued quantities. The appearance of inverse matrices introduces
a high level of non-linearity. 

In \cite{Baggio:2014sna} we provided an explicit independent check that the matrix-valued equations \eqref{reviewaj} are satisfied at tree level 
for any $SU(N)$ ${\cal N}=2$ SCQCD theory. In this work we provide additional non-trivial 
independent evidence for the validity of the \tts equations, by computing the first quantum corrections which arise at 3-loops in perturbation theory.

It is clear that a single datum, like the $S^4$ partition function, is not enough to obtain a full solution 
of the \tts equations in the $SU(N)$ theory with $N>2$. For example, since the scaling dimensions of the fields appearing in \eqref{reviewaj}
are related by an increment of 2, the equations for chiral primaries of even scaling dimension are decoupled
from the equations of the chiral primaries of odd scaling dimension. Hence, separate data are needed to 
solve the \tts equations for the chiral primaries of odd scaling dimension.  
Moreover, even within the sector of even or odd scaling dimensions, the pattern of increasing degeneracies does not admit
an obvious recursive solution as in the simple $SU(2)$ case.
In what follows, we propose a surprising reduction of this problem.

\section{Decoupling the \tts equations}
\label{decoupling}

In this section we set up an ansatz that decouples the \tts equations and allows us to solve them recursively as a set of independent semi-infinite Toda chains.
First, we examine the structure of the chiral ring at tree level. We focus on two natural operators defined on the space of chiral primaries, 
which correspond to the action by the OPE with the $\Delta=2$ chiral primary $\phi_2= \Tr[\varphi^2]$ and the anti-chiral primary 
$\overline{\phi}_2=\Tr[\overline{\varphi}^2]$. We show that these two operators, called $C_2$ and $C_2^\dagger$, 
are the adjoint of one another and satisfy a simple algebra according to which they can be treated as creation and annihilation operators. 
This allow us to decompose the space of chiral primaries into 
representations of this algebra, which are orthogonal with respect to the 2-point functions. In this basis the tree level \tts equations 
explicitly decouple into a set of independent semi-infinite Toda chains. 

At a second stage, we examine how this structure is modified at finite coupling. We argue that the \tts equations \eqref{reviewaj} --- seen abstractly as a  set of coupled differential equations ---  admit a class of solutions, in which different
chiral primaries in the above basis remain orthogonal for all values of the coupling. This class does not necessarily include the most general solution of the 
\tts equations and a priori it is not clear whether it includes the physically relevant solution that we seek in the context of the gauge theory.
Perturbative evidence in favor of the physical relevance of this restricted class is provided in section \ref{perturbation}.

\subsection{The chiral ring at tree level}
\label{c2basis}

The basis of chiral primaries generated by the single-trace operators \eqref{reviewaa} is particularly convenient because the structure 
constants are very simple --- see equation \eqref{reviewae}. Nevertheless, it is straightforward to check that the 2-point functions in this 
basis are \emph{not} diagonal. In turn, this means that the \tts equations in \eqref{reviewaj} do not reduce to simple recursive one-dimensional 
chains of equations, but rather they constitute a set of coupled non-linear partial differential equations where various components of the 2-point 
functions mix nontrivially among themselves. As a result, it makes sense to look for a new basis that diagonalizes the 2-point functions 
while preserving some of the simplicity of \eqref{reviewae}. Since the only structure constants that appear in the \tts equations are 
the ones that involve $\phi_2$, namely $C_{2K}^{L}$, it will suffice to look for a basis where these structure constants remain diagonal, i.e.\
as matrices they have a single non-vanishing element at each row.

Along these lines let us consider first chiral primaries $\phi^{(0)}_K$ with the defining OPE property
\begin{equation}
\label{eq:basicchiralprimary}
\overline{\phi_2}(x) \phi^{(0)}_K(0) = \frac{0}{|x|^{4}} + \ldots
~.
\end{equation}
These are chiral primaries where the most singular term in the OPE with $\overline{\phi_2}$ vanishes.\footnote{It is easy to 
see using $U(1)_R$ conservation and the unitarity bound $\Delta \geq {
|R|\over 2}$ that, as long as $\Delta_K\geq 2$, the most 
singular term on the RHS of \eqref{eq:basicchiralprimary} is of the form  $\frac{\phi(0)}{|x|^{4}}$ where $\phi(0)$ is a chiral primary. 
Of course, in specific cases, such as the $C_2$-primaries that we define above, this term may be absent from the OPE.} 
Henceforth, we will refer to these distinguished chiral primaries as ``$C_2$-primaries''.

We can construct generic chiral primaries by 
acting repeatedly on the $C_2$-primaries with $\phi_2 \propto \Tr\left[\varphi^2\right]$ 
\begin{equation}
\label{eq:basis}
\phi^{(n)}_K \equiv \phi_2^n \phi^{(0)}_K~.
\end{equation}

As an obvious benefit, the structure constants $C_{2K}^L$ are manifestly diagonal in this basis. Hence, for our purposes we would only need to show that the 
2-point functions at tree level are diagonal as well. For starters, let us show that 
\begin{equation}
	\langle \phi^{(m)}_K(x) \, \overline{\phi^{(n)}_L}(0)\rangle = 0~,~~  \mathrm{if} \; m \neq n~.
\end{equation}
In any basis, $\NN=2$ chiral primaries exhibit the OPEs
\begin{align}
	\phi_2(x) \phi_K(0) &= C_{2 K}^L \phi_L(0) + \ldots~,\\
	\overline{\phi_2} (x) \phi_K(0) &= g_{K\overline{R}} C^{*\overline{R}}_{\overline{2}\overline{P}}g^{\overline{P}L} \phi_L(0) \frac{1}{|x|^4} + \ldots
	~.
\end{align}
Hence, as we implied already, there are two natural operators acting on the space of chiral primaries
\begin{align}
	(C_2)^L_K & \equiv C_{2 K}^L~,\\
	\left(C_2^\dagger\right)_K^L & \equiv g_{K\overline{R}} C^{*\overline{R}}_{\overline{2}\overline{P}}g^{\overline{P}L}~.
\end{align}
Put differently, if we have a chiral primary $\phi = v^K \phi_K$, where $v^K$ is an arbitrary vector, then $C_2 \cdot \phi$ is the chiral primary 
that appears in the OPE of $\phi$ with $\phi_2$, and $C_2^\dagger \cdot \phi$ is the chiral primary that appears in the OPE of $\phi$ with 
$\overline{\phi_2}$, that is
\begin{align}
	C_2 \cdot \phi & \equiv v^K C_{2K}^L \phi_L~,\\
	C_2^\dagger \cdot \phi & \equiv v^K g_{K\overline{R}} C^{*\overline{R}}_{\overline{2}\overline{P}}g^{\overline{P}L} \phi_L~.
\end{align}
In particular, the operator $C_2$ raises the $R$-charge of an operator by 4 and the conformal dimension by 2, while $C_2^\dagger$ lowers them by the same amount.

We will now show that $C_2^\dagger$ is the adjoint of $C_2$ with respect to the metric defined by the 2-point functions. Consider the 3-point function
\begin{equation}
	\langle \phi(x_1) \overline{\phi_2} (x_2) \overline{\phi'} (x_3)\rangle = \frac{\alpha}{|x_{12}|^{4}|x_{13}|^{2\Delta-4}}~,
\end{equation}
where $\alpha$ is a constant and we used $\Delta = \Delta' + 2$ due to $R$-charge conservation.
In the limit $x_2 \to x_3 \to 0$, $x_1 \to 1$ we find
\begin{equation}
\label{a1}
	\alpha = \langle \phi(1) \, \overline{C_2\cdot\phi'}(0)\rangle~,
\end{equation}
while in the limit $x_1 \to x_2 \to 1$, $x_3 \to 0$ we find
\begin{equation}
\label{a2}
	\alpha = \langle C_2^\dagger\cdot\phi(1) \, \overline{\phi'}(0)\rangle
	~.
\end{equation}
The combination of equations \eqref{a1}, \eqref{a2} verifies the advertised statement: 
$C_2^\dagger$ is indeed the adjoint of $C_2$ with respect to the metric induced by the 2-point functions.

There is a second crucial property of the operators $C_2$, $C_2^\dagger$. Their commutator acts as follows
\begin{align}
\label{eq:commutator}
[C_2,C_2^\dagger]\cdot \phi & = v^K \left( g_{K\overline{R}} C^{*\overline{R}}_{\overline{2}\overline{P}}g^{\overline{P}L} C_{2L}^Q 
- C_{2K}^L g_{L\overline{R}} C^{*\overline{R}}_{\overline{2}\overline{P}}g^{\overline{P}Q}\right)\phi_Q \\
& \equiv - v^K [C_2,\overline{C_2}]_K^Q\, \phi_Q~.
\end{align}
The combination $[C_2,\overline{C_2}]_K^Q$ satisfies at tree level a nice combinatorial identity that was proven in appendix C 
of \cite{Baggio:2014ioa}
\begin{equation}
[C_2,\overline{C_2}]_K^L = g_{2\overline{2}} \delta_K^L \left(1+\frac{R}{\mathrm{dim}\,\mathcal{G}}\right)~.
\end{equation}
Hence, when we plug this identity into \eqref{eq:commutator}, we find
\begin{equation}
\label{c2algebra}
	[C_2,C_2^\dagger]\cdot \phi = -g_{2\overline{2}} \left(1+\frac{R}{\mathrm{dim}\,\mathcal{G}}\right) \phi~.
\end{equation}
This means that we can regard $C_2$ and $C_2^\dagger$ as creation and annihilation operators respectively. In other words, 
we can decompose the space of chiral primaries in terms of representations of this algebra. We start from 
`highest weight chiral primaries' annihilated by $C_2^\dagger$ (the $C_2$-primaries),
and we build the space of states by acting with $C_2$ (multiplying by $\phi_2$, as in \eqref{eq:basis}). 
Then, the resulting representations must necessarily be orthogonal, i.e.\ they diagonalize the 2-point functions. 
At this stage this is easy to verify directly if, say, $m<n$ and $C_2^\dagger \cdot \phi^{(0)} = 0$. Indeed,
\begin{equation}
	\langle C_2^m \cdot \phi^{(0)}(x)\, \overline{C_2^n\cdot  \phi'}(0) \rangle 
	= \langle (C_2^\dagger)^n C_2^m \cdot \phi^{(0)}(x) \,\overline{\phi'}(0) \rangle = 0 
	~,
\end{equation}
where in the last step we used repeatedly the commutator $[C_2,C_2^\dagger]$.

In the special case $m = n$, i.e.\ when two chiral primaries $\phi$, $\phi'$ are degenerate, we obtain similarly the identity
\begin{equation}
	\langle C_2^m \cdot \phi(x)\, \overline{C_2^m \cdot \phi'}(0) \rangle \propto g_{2\overline{2}}^m \langle \phi(x) \overline{\phi'}(0) \rangle~,
\end{equation}
Consequently, if there is a degeneracy in the spectrum of $C_2$-primaries, we can choose any orthogonal combination 
and the orthogonality will be preserved by the $C_2$-descendants. As a result, the basis built on $C_2$-primaries that is singled out
in this section is not unique. At tree level any orthogonal combination of degenerate $C_2$-primaries is equally acceptable for our
purposes. It is nevertheless useful to keep this freedom in mind in the context of finite coupling effects, which will be discussed shortly.

\subsection{\tts equations at tree level}
\label{tree}

It is rather straightforward to show that, in the basis defined in the previous section, the tree level \tts equations decouple into a collection of 
one-dimensional semi-infinite Toda chains. Indeed, starting with a given arbitrary $C_2$-primary $\phi^{(0)}$, let us consider the subsequence of 
chiral primaries $\phi^{(n)} = \phi_2^n \phi^{(0)}$. Since the 2-point functions are diagonal at tree level, we can focus on the components
\begin{equation}
G_{2n} \equiv \langle \phi^{(n)}(1) \overline{\phi^{(n)}}(0)\rangle~.
\end{equation}
Inserting the results of the previous subsection into the tree level version of \eqref{reviewaj}, it is easy to see that the $G_{2n}$ satisfy
\begin{equation}
\label{eq:gentts}
\partial_{\barM \tau} \partial_\tau \log G_{2n} = \frac{G_{2n+2}}{G_{2n}} - \frac{G_{2n}}{G_{2n-2}} - g_2~,
\end{equation}
which is very similar to the one-dimensional chain \eqref{reviewaka} of the $SU(2)$ case (and can be recast as the semi-infinite Toda chain \eqref{reviewal}).

At tree level it is not hard to solve \eqref{eq:gentts} explicitly in closed form. Let us assume that the $C_2$-primary $\phi^{(0)}$ has scaling dimension $\Delta_0$.
Then, the generic chiral primary $\phi^{(n)} = \phi_2^n \phi^{(0)}$ has scaling dimension $\Delta_0+2n$, and at tree level
\beq
\label{treeaa}
G_{2n} = \frac{1}{\left( {\rm Im}\tau \right)^{\Delta_0+2n}} \tilde G_{2n}
~,
\eeq
where $\tilde G_{2n}$ are $(\tau,\barM\tau)$-independent constants determined solely by group-theoretical Wick contractions.
Implementing \eqref{treeaa} equation \eqref{eq:gentts} becomes
\beq
\label{treeab}
\frac{\Delta_0+2n}{4} = \frac{\tilde G_{2n+2}}{\tilde G_{2n}} - \frac{\tilde G_{2n}}{\tilde G_{2n-2}} - \tilde g_2
~.
\eeq
Moreover, in our conventions
\beq
\label{treeac}
\tilde g_2 = \frac{N^2-1}{8}
~.
\eeq
Consequently, solving \eqref{treeab} we obtain 
\beq
\label{treead}
\tilde G_{2n} = \frac{\tilde G_0}{4^n} 
\prod_{\ell=0}^{n-1} 
\left[ \frac{4 \tilde G_2}{\tilde G_0} + \ell \left( \frac{N^2-1}{2} + \Delta_0 +1\right) + \ell^2 \right]
\eeq
in terms of the numerical 2-point function coefficients $\tilde G_0$, $\tilde G_2$ for the correlators $\langle \phi^{(n)}(1) \overline{\phi^{(n)}}(0)\rangle$
with $n=0,1$.

In the special case, where $\phi^{(0)} = \mathbf{1}$ (the identity operator), the sequence $\phi^{(n)} = \phi_2^n$ 
is comprised of the same type of operators that constitute the $\NN=2$ chiral ring in the $SU(2)$ case. 
For those operators we have $G_{2n} = g_{2n}$, and equation \eqref{eq:gentts} 
is exactly the same semi-infinite Toda chain that was encountered in the $SU(2)$ case \eqref{reviewaka}.
In this situation the solution \eqref{treead} simplifies to
\beq
\label{treeaf}
\tilde g_{2n} = \frac{n!}{4^n} \left( \frac{N^2-1}{2}\right)_n 
~,
\ee
where $(x)_n$ is the Pochhammer symbol
\beq
\label{treeae}
(x)_n = x(x+1)\cdots (x+n-1)
~.
\eeq
This relation was noticed empirically and conjectured to hold for the general $SU(N)$ theory in \cite{Baggio:2014ioa}.
Amusingly, a similar relation has been proven with direct methods some time ago in appendix A.4 of Ref.\ \cite{Brown:2006zk} for the $U(N)$ $\NN=4$ SYM theory.
The $SU(N)$ and $U(N)$ formulae are identical with the suggestive substitution of $N^2-1$ with $N^2$ (the dimension of the gauge group) 
inside the Pochhammer symbol \eqref{treeaf}. The formula \eqref{treead} is an interesting generalization to arbitrary $\NN=2$ chiral primary operators.
It is equally applicable to chiral primary operators in $SU(N)$ $\NN=4$ SYM theory, where the tree level 2- and 3-point functions do not 
receive quantum corrections.

\subsection{\tts equations at finite coupling: a no-mixing ansatz}
\label{finitecoupling}

Having shown at tree level that the \tts equations decouple into a sequence of independent Toda chains, it is natural to ask if a similar decoupling
continues to hold when quantum corrections are taken into account. A sufficient condition for this effect is the requirement that the full non-perturbative
2-point functions remain diagonal in at least one of the bases constructed in the previous subsections, call it $\hat \phi_K$ 
(recall that the previous tree level construction of the bases based on $C_2$-primaries was not unique).
This requirement is mathematically consistent from the point of view of the \tts equations.  Notice that what we are postulating here is essentially the ability to diagonalise the exact 2-point functions within the holomorphic gauge. 

From the point of view of the gauge theory the no-mixing condition postulated by this ansatz appears to be a new non-renormalization theorem 
in a four-dimensional $\NN=2$ theory. Notice that unlike the non-renormalization theorem in $\NN=4$ SYM \cite{Lee:1998bxa,D'Hoker:1998tz,D'Hoker:1999ea,Intriligator:1998ig,Intriligator:1999ff,Eden:1999gh,Petkou:1999fv,Howe:1999hz,Heslop:2001gp,Baggio:2012rr}, this theorem would 
not fix completely the moduli-dependence
of correlation functions in the chiral ring. So far we have not been able to prove it using superconformal Ward identities. If true, this theorem would lead to several non-trivial consequences, which are discussed in detail in the next section. For instance, it would imply geometrically
that the gauge connection of the holomorphic chiral primary vector bundles on the $\NN=2$ superconformal manifold are reducible.

In appendix \ref{explicit} we also formulate the no-mixing condition in terms of the original multi-trace basis $\phi_K$ of equation \eqref{reviewaaa}. 
In that basis, the non-renormalization condition translates into a statement about the coupling constant independence of appropriate ratios of 
2-point functions. 

In section \ref{perturbation} we put the above ansatz to the test in perturbation theory by computing the first non-trivial quantum corrections to 
several 2-point functions of chiral primary operators in $SU(3)$ and $SU(4)$ 
SCQCD. We proceed as high in scaling dimension and gauge group rank as possible, given our current computational limitations with the complicated 
combinatoric structures at 3-loops. In all cases, we verify the no-mixing ansatz: the 2-point functions remain diagonal, and the decoupled Toda equations
\eqref{eq:gentts} are explicitly verified. This evidence seems to be highly suggestive.

\section{Implications of the decoupling}
\label{implications}

In this section, we explore some of the implications of the conjectured no-mixing condition and the related decoupling of the \tts equations. 
First, we show that, similar to the $SU(2)$ case \cite{Baggio:2014ioa}, we can use the decoupled Toda chains 
to determine exact correlation functions in the $\NN=2$ chiral ring of the $SU(N)$ theory from a single datum in each decoupled subsector.
In particular, we show that the subsequence based on the identity operator
is exactly solvable using current knowledge from supersymmetric localization. 
With the same data we also obtain predictions for the exact form of certain extremal correlators
that involve only single-trace operators, e.g.\ certain single-trace 3-point functions.

Second, we examine the geometric interpretation of this decoupling 
and show that it implies that the holonomy group on the space of chiral primaries is restricted --- assuming full decoupling 
the holonomy group is a product of abelian factors. 

Finally, we present other implications of the no-mixing ansatz on general (not necessarily extremal) integrated correlation functions.

\subsection{Proposed recursive solution of the \tts equations at finite coupling}

We begin by rewriting \eqref{eq:gentts} in the recursive form
\begin{equation}
\label{recurseaa}
	G_{2n+2} = G_{2n} \partial_{\tau}\partial_{\overline{\tau}}\log\,G_{2n} + \frac{G_{2n}^2}{G_{2n-2}} + G_{2n}g_2~.
\end{equation}
Assuming the no-mixing ansatz of subsection \ref{finitecoupling} this is now an equation that holds non-perturbatively in 
the $SU(N)$ theory at finite coupling. The solution is determined recursively from the 2-point function of the $C_2$-primary 
operator $\phi^{(0)}$ under consideration
\begin{equation}
	G_{0} = \langle \phi^{(0)}(1)\,\overline{\phi^{(0)}}(0)\rangle
\end{equation}
and the 2-point function of the unique $\Delta=2$ chiral primary $\phi_2$
\begin{equation}
	g_2 = \langle \phi_2(1)\, \overline{\phi_2}(0)\rangle
~,
\end{equation}
which is, up to a convention-dependent numerical coefficient, the Zamolodchikov metric. Different choices of the operator $\phi^{(0)}$ sample different
subsectors of the $\NN=2$ chiral ring and correspond to different solutions of the recursive equations \eqref{recurseaa}.

As we noted already in section \ref{review}, the Zamolodchikov metric can be obtained from the $S^4$ partition function using supersymmetric localization 
\cite{Gerchkovitz:2014gta,Pestun:2007rz}. In the $SU(N)$ $\NN=2$ SCQCD theory the $S^4$ partition function can be written as an $(N-1)$-dimensional
ordinary integral. We refer the reader to the original references for explicit formulae. 

To the best of our knowledge, it is not currently known how to compute the general $G_{0}$ exactly as a function of the moduli for arbitrary $C_2$-primaries. 
A notable exception is the main subsequence defined by the identity operator, $\phi^{(0)} = \mathbf{1}$. As we pointed out already in subsection \ref{tree}, 
in this case the 2-point functions $G_{2n} \equiv g_{2n}$ satisfy equation \eqref{reviewaka}, so the analysis of \cite{Baggio:2014ioa,Baggio:2014sna} can be 
repeated almost without changes. The only difference is that the starting point $g_2$ must be computed from the $S^4$ partition function 
for the gauge group $SU(N)$ instead of $SU(2)$.

\subsection{Single-trace extremal correlation functions and large-$N$ limits}

According to our conjecture, current knowledge of the Zamolodchikov metric also gives exact predictions for certain 
extremal correlation functions that involve only \emph{single-trace} chiral primaries. Correlation functions of single-trace operators 
have an obvious interest in large-$N$ limits.

As an illustrating example, let us consider first such a 3-point function of single-trace operators in the $SU(4)$ theory. 
It will be shown in the next section that the only $C_2$-primary at scaling dimension $\Delta=4$ is
\begin{equation}
	\phi^{(0)}_4 = \Tr[\varphi^4] - \frac{29}{68} \Tr[\varphi^2]^2~.
\end{equation}
Our conjecture implies in particular that
\begin{equation}
\label{singleaa}
	\langle\phi_2(\infty) \phi_2(1) \overline{\phi^{(0)}_4}(0) \rangle = 0~.
\end{equation}
Moreover, we know that
\begin{equation}
	\langle\phi_2(\infty) \phi_2(1) \overline{\phi_2^2}(0) \rangle = C_{\phi_2 \phi_2}^{\phi_2^2} \langle \phi_2^2(1) \overline{\phi_2^2}(0)\rangle
	= g_4
	~,
\end{equation}
with $g_4$ being determined from $g_2$ and the Toda equation as 
\beq
\label{singleab}
g_4= g_2 \, \partial_{\barM \tau} \partial_\tau \log g_2 + 2\, g_{2}^2
~.
\eeq 
Together with equation \eqref{singleaa} we deduce that
\begin{equation}
\label{singleac}
	\langle\Tr[\varphi^2](\infty)\, \Tr[\varphi^2](1)\, \Tr[\overline{\varphi}^4](0) \rangle = \frac{29}{68} g_4~.
\end{equation}

Notice that \eqref{singleac} is an exact expression for a 3-point function in a specific set of normalization conventions, where the 2-point functions
$\langle \Tr[\varphi^2](1)\, \Tr[\overline{\varphi}^2](0)\rangle$, $\langle \Tr[\varphi^4](1)\, \Tr[\overline{\varphi}^4](0)\rangle$ are fixed.
Although it is known at the moment how to compute exactly the first of these 2-point functions using localization, it is not known how to compute 
the second. 

The result \eqref{singleac} has the following straightforward generalization. From the analysis of Ref.\ \cite{Baggio:2014ioa} we can deduce, using 
superconformal Ward identities, the identity
\beq
\label{singlead}
\langle \Tr[\varphi^2] (x_1) \Tr[\varphi^2] (x_2) \cdots \Tr[\varphi^2](x_{k}) \Tr[\overline{\varphi}^{2k}](\infty) \rangle = 
\langle \Tr[\varphi^2]^k(0) \Tr[\varphi^{2k}](\infty) \rangle
~.
\eeq
The extremal $(k+1)$-point function in question is independent of the insertions of the operators and equal to a 2-point function for two operators at
scaling dimension $2k$. Following an argument similar to the one of the previous paragraphs, or equivalently the non-renormalization identities of appendix \ref{explicit},
we obtain
\beq
\label{singleae}
\langle \Tr[\varphi^2]^k(0) \Tr[\varphi^{2k}](\infty) \rangle 
= \frac{\langle \Tr[\varphi^2]^k(0) \Tr[\varphi^{2k}](\infty) \rangle_{tree}}
{\langle \Tr[\varphi^2]^k(0) \Tr[\varphi^2]^k(\infty) \rangle_{tree}}\, g_{2n}
~,
\eeq
where the prefactor is evaluated at tree level, and $g_{2n}$ is determined from the $S^4$ partition function and the chain \eqref{reviewaka}.
 
It would be interesting to study the behavior of such single-trace correlators further in the large-$N$ limit and explore possible implications in 
related applications of the AdS/CFT correspondence. We note in passing that, as a simple check of our formalism and the 
\tts equations \eqref{reviewaj} in the large-$N$ limit, one can easily verify that large-$N$ factorization is an automatic solution of the \tts equations. 

More along the lines of the large-$N$ limit, the recent work \cite{Louis:2015dca} studied in supergravity the structure of the moduli space of
certain supersymmetric AdS$_5$ vacua, which have the right amount of supersymmetry to be the  holographic duals of 4d 
${\cal N}=2$ SCFTs. If these theories have a holographic dual, then the moduli space of vacua in supergravity correspond to the 
conformal manifold of the dual SCFT.  It would be interesting to investigate the form
of the Zamolodchikov metric in the large-$N$ limit directly from the gauge theory and to compare it with the supergravity 
results of \cite{Louis:2015dca}.

\subsection{Reducible chiral primary bundles}

Another consequence of the existence of a holomorphic basis $\hat \phi_K$ that diagonalizes the 2-point functions non-perturbatively 
is that the vector bundles of chiral primaries are \emph{reducible}. It is readily seen from equations \eqref{reviewaia}-\eqref{reviewaib}, 
that the connection $\hat A$ is diagonal in the basis of the operators $\hat \phi_K$. Consequently, if there are $D$ chiral primaries of scaling dimension $\Delta$
(at arbitrary $\Delta$), the holonomy will be restricted to the subgroup $U(1)^D \subset U(D)$. A reducible holonomy 
is a non-trivial condition for the geometry of the chiral primary vector bundles over the superconformal manifold.

The strong version of the no-mixing conjecture proposed in subsection \ref{finitecoupling} states that the 2-point functions are fully diagonalizable
non-perturbatively in a holomorphic basis, hence the connection and the associated holonomy are fully reducible. Notice that full reducibility
is consistent with the operator product structure 
\beq
\label{reducibleaa}
C~:~ \VV_\Delta \times \VV_{\Delta'} ~\to ~ \VV_{\Delta+\Delta'}
\eeq
that allows us to multiply sections from two chiral primary vector bundles to obtain a section on a third chiral primary vector bundle at the sum of
scaling dimensions.

Currently we have not excluded the possibility of a consistent weaker version of the no-mixing conjecture, where the holonomy is 
partially reducible to a subgroup that is a product of abelian and non-abelian factors. In the next section, where we provide direct evidence
for decoupling in perturbation theory, we verify that 2-point functions
\begin{equation}
	\langle \phi^{(m)}(x)\,\overline{\phi^{(n)}}(0)\rangle = 0~, ~~ \mathrm{with}~ m \neq n
\end{equation}
do not mix at the quantum level. In all the cases that we have analyzed so far, the degenerate operators are $C_2$-descendants of primaries 
at different scaling dimensions. Interesting subtleties, with potential non-abelian holonomies, could seemingly appear in situations with more than one 
degenerate $C_2$-primary operators. Recall that this was precisely the origin of the non-uniqueness of the basis constructed from the $C_2$-algebra
in section \ref{c2basis}.

For example, if $N \geq 6$, the $\Delta=6$ spectrum includes the operators
\begin{equation}
	\Tr\left[\varphi^6\right]\, , ~~ \Tr\left[\varphi^3\right]^2 \, ,~~ \Tr\left[\varphi^4\right] \Tr \left[ \varphi^2 \right] \, , ~~ \Tr\left[\varphi^2\right]^3 
	~.
\end{equation}
It is clear that we can build two independent $C_2$-primary combinations out of the operators in this list. At the moment, we cannot exclude
the possibility that there is no constant linear combination of these two $C_2$-primaries that keeps them orthogonal at finite coupling.
Verifying what actually happens would require a perturbative computation at more than 3 loops, which lies beyond our current computational power.
Therefore, we cannot currently provide decisive evidence that favors a $U(1)^4$ holonomy compared to a $U(1)^2 \times U(2)$ holonomy in this sector.

\subsection{Other implications}

The reducibility of the connection has further implications, even for non-extremal correlation functions in the $\NN=2$ chiral ring.
Consider the general $(n+\barM n)$-point function in the $\NN=2$ chiral ring
in the diagonal hatted basis $\hat\phi_K$
\beq
\label{impliba}
\AA_{K_1 \cdots K_n \barM L_1 \cdots \barM L_{\barM n}}
=\langle \hat \phi_{K_1}(x_1) \cdots \hat \phi_{K_n}(x_n) \overline{\hat \phi_{L_1}}(y_1) \cdots \overline{\hat \phi_{L_{\barM n}}}(y_{\barM n}) \rangle
\eeq
where the total R-charge of the insertions vanishes. The covariant derivative of this correlation function with respect to the
complexified gauge coupling $\tau$ expresses by definition \cite{Papadodimas:2009eu,Baggio:2014ioa} the renormalized
integrated $(n+\barM n+1)$-point function
\bea
\label{implibb}
\hat \nabla_\tau \AA_{K_1 \cdots K_n \barM L_1 \cdots \barM L_{\barM n}}
&=& \Big \langle
\int d^4 z\, \OO_\tau(z) 
\hat \phi_{K_1}(x_1) \cdots \hat \phi_{K_n}(x_n) \overline{\hat \phi_{L_1}}(y_1) \cdots \overline{\hat \phi_{L_{\barM n}}}(y_{\barM n}) 
\Big \rangle_{renormalized}
\nonumber\\
&=& \partial_\tau  \AA_{K_1 \cdots K_n \barM L_1 \cdots \barM L_{\barM n}}
-\sum_{i=1}^n \left( \hat A_\tau \right)_{K_i}^M \,  \AA_{K_1 \cdots K_{i-1} M\cdots K_n \barM L_1 \cdots \barM L_{\barM n}}
\nonumber\\
&=& \partial_\tau  \AA_{K_1 \cdots K_n \barM L_1 \cdots \barM L_{\barM n}}
- \left( \sum_{i=1}^n \left( \hat A_\tau \right)_{K_i}^{K_i} \right)  \AA_{K_1 \cdots K_n \barM L_1 \cdots \barM L_{\barM n}}
\eea
where in the last step we assumed the full reducibility of the connection.

A characteristic example of the general relation \eqref{implibb} is the case of the covariant derivative of the 3-point function
\beq
\label{implibc}
\hat C_{2K\barM L} = \hat C_{2K}^M \hat g_{M\barM L} = \hat g_{K+2,\barM L} = \hat g_{K+2,\barM K+\barM 2} \, \delta_{\barM K+\barM 2, \barM L}
~.
\eeq
Direct computation of the RHS of equation \eqref{implibb} in this case implies the vanishing of the integrated 4-point function
\beq
\label{implibd}
\Big \langle
\int d^4 z\, \OO_\tau(z) \, 
\phi_2(x_1) \hat \phi_{K}(x_2) \overline{\hat \phi_{L}}(y)  \Big \rangle_{renormalized} =0~, ~~ \barM L\neq \barM K+\barM 2
~.
\eeq
Since there is no obvious symmetry reason for this identity, it would be interesting to obtain it with an independent derivation.
We suspect that such a derivation might be a useful step towards the ultimate proof of the no-mixing conjecture.

\section{Checks in perturbation theory}
\label{perturbation}

In this section we compute the first non-trivial quantum corrections to the 2-point functions of  chiral primaries in certain 
examples in $SU(N)$ SCQCD. The first non-trivial correction appears diagrammatically at 3-loops.
In all examples we find evidence that the connection on the space of chiral primaries is indeed reducible in accordance with the no-mixing
proposal of section \ref{finitecoupling}.

More specifically, using the general 3-loop perturbative formula of appendix \ref{pert}, \eqref{pertbg},
we compute the perturbative matrix of 2-point functions up to conformal dimension $\Delta=8$ for $SU(3)$ and $\Delta=6$ for $SU(4)$.
The explicit computation was performed with Mathematica. We report only these cases at this stage, because as we increase the rank $N$ and the 
scaling dimension $\Delta$ of the operators, the combinatorics of the general formula \eqref{pertbg} quickly render the computation slow and 
impractical.

\subsection{$SU(3)$ SCQCD up to $\Delta=8$}

We begin with the analysis of 2-point functions in the $SU(3)$ theory. In this case, the $\NN=2$ chiral ring is generated by the chiral primaries
\begin{equation}
\Tr[\varphi^2]~, ~~ \Tr[\varphi^3]~.
\end{equation}
The first scaling dimension with non-trivial degeneracy is $\Delta=6$, where we have the operators
\begin{equation}
\Tr[\varphi^2]^3~, ~~ \Tr[\varphi^3]^2~.
\end{equation}
Notice that in order to determine whether or not the $tt^*$ equations decouple, we need to study 2-point functions up to level 8.

Applying the formulae and normalization conventions of appendix \ref{pert}, we find the following results
\begin{align}
G_2 & = \left(\frac{\gym^2}{16\pi}\right)^2\left(16 - \frac{45 \,\zeta(3)}{2\pi^4} \gym^4\right)~,\\[5px]
G_3 & = \left(\frac{\gym^2}{16\pi}\right)^3\left(40 - \frac{135 \,\zeta(3)}{2\pi^4} \gym^4\right)~,\\[5px]
G_4 & = \left(\frac{\gym^2}{16\pi}\right)^4\left(640 - \frac{2160 \,\zeta(3)}{\pi^4} \gym^4\right)~,\\[5px]
G_5 & = \left(\frac{\gym^2}{16\pi}\right)^5\left(1120 - \frac{4410 \,\zeta(3)}{\pi^4} \gym^4\right)~,\\[5px]
G_6 & = \left(\frac{\gym^2}{16\pi}\right)^6
\begin{pmatrix}
46080 - \frac{272160 \,\zeta(3)}{\pi^4} \gym^4 &~ 1920 - \frac{11340 \,\zeta(3)}{\pi^4} \gym^4 \\[5px]
1920 - \frac{11340 \,\zeta(3)}{\pi^4} \gym^4 &~ 6800 - \frac{57645 \,\zeta(3)}{2\pi^4} \gym^4
\end{pmatrix}~,\\[5px]
G_7 & = \left(\frac{\gym^2}{16\pi}\right)^7\left(71680 - \frac{483840 \,\zeta(3)}{\pi^4} \gym^4\right)~,\\[5px]
G_8 & = \left(\frac{\gym^2}{16\pi}\right)^8
\begin{pmatrix}
5160960 - \frac{46448640 \,\zeta(3)}{\pi^4} \gym^4 &~ 215040 - \frac{1935360 \,\zeta(3)}{\pi^4} \gym^4 \\[5px]
215040 - \frac{1935360 \,\zeta(3)}{\pi^4} \gym^4 &~ 277760 - \frac{2046240 \,\zeta(3)}{\pi^4} \gym^4
\end{pmatrix}~.
\end{align}

The $2\times 2$ matrix $G_6$ is written in the basis $\Tr[\varphi^2]^3, \Tr[\varphi^3]^2$, 
while $G_8$ is written in the basis $\Tr[\varphi^2]^4, \Tr[\varphi^2]\Tr[\varphi^3]^2$.
It is manifest that this basis does not diagonalize the 2-point functions, not even at tree-level. 
As explained in previous sections, we can diagonalize the 2-point functions by constructing the $C_2$-primaries. $\Tr[\varphi^3]^2$, in particular, is not a $C_2$-primary, as can be easily seen from the tree-level OPEs
\begin{align}
\Tr[\overline{\varphi}^2](x) \Tr[\varphi^2]^3(0) & \approx \frac{9 \gym^4}{32\pi^2|x|^4} \Tr[\varphi^2]^2 (0) +\ldots~,\\[5px]
\Tr[\overline{\varphi}^2](x) \Tr[\varphi^3]^2(0) & \approx \frac{3 \gym^4}{256\pi^2|x|^4} \Tr[\varphi^2]^2(0) +\ldots~.
\end{align}
It is then easy to take an appropriate linear combination of the two chiral primaries of dimension 6 that is annihilated by $C_2^\dagger$. 
The appropriate bases at scaling dimensions 6 and 8 are then given by the operators
\begin{align}
\phi_{6} & = \Tr[\varphi^2]^3~, & \phi_{6'} & = \Tr[\varphi^3]^2 - \frac{1}{24} \Tr[\varphi^2]^3~,\\
\phi_{8} & = \phi_2 \phi_6~, & \phi_{8'} & = \phi_2 \phi_{6'}~,
\end{align}
where it is easily checked that $C_2^\dagger \cdot \phi_{6'} = 0$.
In the new basis, the 2-point functions become diagonal even when we include the first non-trivial quantum corrections
\begin{align}
G'_6 & = \left(\frac{\gym^2}{16\pi}\right)^6
\begin{pmatrix}
46080 - \frac{272160 \,\zeta(3)}{\pi^4} \gym^4 & 0 \\[5px]
0 &~ 6720 - \frac{28350 \,\zeta(3)}{\pi^4} \gym^4
\end{pmatrix}~,\\[5px]
G'_8 & = \left(\frac{\gym^2}{16\pi}\right)^8
\begin{pmatrix}
5160960 - \frac{46448640 \,\zeta(3)}{\pi^4} \gym^4 & 0 \\[5px]
0 &~ 268800 - \frac{1965600 \,\zeta(3)}{\pi^4} \gym^4
\end{pmatrix}
\end{align}
verifying at this order the no-mixing conjecture.

It is also easy to check that the correlators satisfy the appropriate Toda chains \eqref{eq:gentts}, as explained in the previous sections.

\subsection{$SU(4)$ SCQCD up to $\Delta=6$}

In this section we study correlation functions in the $SU(4)$ theory. In this case the $\NN=2$ chiral ring is generated by the three chiral primaries
\begin{equation}
\Tr[\varphi^2]~, ~~ \Tr[\varphi^3]~, ~~ \Tr[\varphi^4]~.
\end{equation}
Consequently, the spectrum is already degenerate at $\Delta=4$, where we have the two degenerate operators
\begin{equation}
\Tr[\varphi^2]^2~, ~~ \Tr[\varphi^4]~.
\end{equation}
At $\Delta=6$, we have an additional degeneracy compared to the $SU(3)$ case, as we have the three independent operators
\begin{equation}
\Tr[\varphi^2]^3~, ~~ \Tr[\varphi^2]\Tr[\varphi^4]~, ~~ \Tr[\varphi^3]\Tr[\varphi^3]~.
\end{equation}

Applying the formulae of appendix \ref{pert}, we find the 2-point functions
\begin{align}
G_2 & = \left(\frac{\gym^2}{16\pi}\right)^2\left(30 - \frac{2295 \,\zeta(3)}{32\pi^4} \gym^4\right)~,\\[5px]
G_3 & = \left(\frac{\gym^2}{16\pi}\right)^3\left(135 - \frac{23085 \,\zeta(3)}{64\pi^4} \gym^4\right)~,\\[5px]
G_4 & =
\begin{pmatrix}
2040 - \frac{43605 \,\zeta(3)}{4\pi^4} \gym^4 &~ 870 - \frac{74385 \,\zeta(3)}{16\pi^4} \gym^4 \\[5px]
870 - \frac{74385 \,\zeta(3)}{16\pi^4} \gym^4 &~ \frac{1335}{2} - \frac{198045 \,\zeta(3)}{64\pi^4} \gym^4
\end{pmatrix}~,\\[5px]
G_5 & = \left(\frac{\gym^2}{16\pi}\right)^5\left(5670 - \frac{535815 \,\zeta(3)}{16\pi^4} \gym^4\right)~,\\[5px]
G_6 & = \left(\frac{\gym^2}{16\pi}\right)^6
\begin{pmatrix}
232560 - \frac{8241345 \,\zeta(3)}{4\pi^4} \gym^4 &~ 99180 - \frac{14058765 \,\zeta(3)}{16\pi^4} \gym^4 &~ 6480 - \frac{229635 \,\zeta(3)}{4\pi^4} \gym^4\\[5px]
99180 - \frac{14058765 \,\zeta(3)}{16\pi^4} \gym^4 &~ 55935 - \frac{30324105 \,\zeta(3)}{64\pi^4} \gym^4 &~ 8100 - \frac{1012095 \,\zeta(3)}{16\pi^4} \gym^4\\[5px]
6480 - \frac{229635 \,\zeta(3)}{4\pi^4} \gym^4 &~ 8100 - \frac{1012095 \,\zeta(3)}{16\pi^4} \gym^4 &~ 58320 - \frac{1454355 \,\zeta(3)}{4\pi^4} \gym^4
\end{pmatrix}.
\end{align}
As before, we can find a constant linear rotation that diagonalizes the 2-point functions at tree level by finding the appropriate $C_2$-primary 
combinations. The new basis is given by the operators
\begin{align}
\phi_{4} & = \Tr[\varphi^2]^2~, & \phi_{4'} & = \Tr[\varphi^4] - \frac{29}{68} \Tr[\varphi^2]^2~,
\end{align}
\begin{align}
\phi_{6} & = \phi_2 \phi_4~, & \phi_{6'} & = \phi_2 \phi_{6'}~, 
& \phi_{6''} & = \Tr[\varphi^6] - \frac{9}{23} \Tr[\varphi^2]\Tr[\varphi^4] + \frac{243}{1748}\Tr[\varphi^2]^3
\end{align}
and the corresponding 2-point functions are given by
\begin{align}
G'_4 & = \left(\frac{\gym^2}{16\pi}\right)^4
\begin{pmatrix}
2040 - \frac{43605 \,\zeta(3)}{4\pi^4} \gym^4 & 0 \\[5px]
0 & \frac{5040}{17} - \frac{18900 \,\zeta(3)}{17\pi^4} \gym^4
\end{pmatrix}~,\\[5px]
G'_6 & = \left(\frac{\gym^2}{16\pi}\right)^8
\begin{pmatrix}
232560 - \frac{8241345 \,\zeta(3)}{4\pi^4} \gym^4 & 0 & 0\\[5px]
0 & \frac{231840}{17} - \frac{3368925 \,\zeta(3)}{34\pi^4} \gym^4 & 0\\[5px]
0 & 0 & \frac{24494400}{437} - \frac{151559100 \,\zeta(3)}{437\pi^4} \gym^4
\end{pmatrix}~,
\end{align}
Once again the no-mixing conjecture is verified.

It is also easy to verify that the diagonal components of the above 2-point functions obey \eqref{eq:gentts}.

\section{Outlook}
\label{outlook}

The observations in this paper suggest the existence of a new interesting class of non-renormalization theorems in 
four-dimensional $\NN=2$ superconformal field theories. It would be important to prove these theorems in 
the $SU(N)$ $\NN=2$ SCQCD theories, and to clarify whether the holonomy of the chiral primary vector bundles is fully 
or partially reducible. 

We emphasized that full reducibility is a consistent ansatz from the point of view of the \tts equations, which reduces them
to an independent set of semi-infinite Toda chains. The non-perturbative solution of the 2-point functions in each of these chains 
requires a single external datum. It would be interesting to explore techniques that will allow the exact computation of
these data generalizing the success of supersymmetric localization on the four-sphere for the Zamolodchikov metric.

We would also like to highlight the efficiency of our results already at tree level. The tree level formulae derived in this paper
are also applicable in the same form in the context of chiral primaries in $\NN=4$ SYM theory.

In conclusion, in this paper we have seen that the study of the \tts equations is a powerful guide towards new exact results in four-dimensional
quantum field theories. It would be extremely interesting to study the solution of the \tts equations in more general classes of
$\NN=2$ superconformal field theories, and to examine the possibility of more general non-renormalization theorems in 
$\NN=2$ theories. At face value, the appearance of such theorems in $\NN=2$ theories is rather unexpected. Perhaps
there are similar surprises in $\NN=1$ theories as well. It would be interesting to explore this possibility.

\section*{Acknowledgments}

\noindent We would like to thank Jan de Boer, Jaume Gomis, Robert de Mello Koch, Zohar Komargodski, 
Jan Louis, Wolfgang Lerche, Sanjaye Ramgoolam, Hagen Triendl 
and Cumrun Vafa for useful discussions. A preliminary version of the results in this work were reported at Strings 2015, 
the 8th Crete Regional Meeting on String Theory, and the Integrability in Gauge and String Theory (IGST) 2015. V.N. would like to
thank many of the participants of these conferences for useful comments and conversations.
We used JaxoDraw \cite{Binosi:2003yf,Binosi:2008ig} to draw all the Feynman diagrams in this paper, and 
Mathematica to perform the explicit combinatorics in section \ref{perturbation}. K.P. would like to thank the University of Crete for hospitality, where
part of this work was completed.
The work of V.N. was supported in part by European Union's Seventh Framework
Programme under grant agreements (FP7-REGPOT-2012-2013-1) no 316165,
PIF-GA-2011-300984, the EU program ``Thales'' MIS 375734 and was also co-financed
by the European Union (European Social Fund, ESF) and Greek national funds through
the Operational Program ``Education and Lifelong Learning'' of the National Strategic
Reference Framework (NSRF) under ``Funding of proposals that have received a positive
evaluation in the 3rd and 4th Call of ERC Grant Schemes''. K.P. would like to thank the 
Royal Netherlands Academy of Sciences (KNAW).

\appendix
\addtocontents{toc}{\protect\setcounter{tocdepth}{1}}
\addtocontents{lof}{\protect\setcounter{tocdepth}{1}}

\section{Perturbative 2- and 3-point functions in $SU(N)$ $\NN=2$ SCQCD theory}
\label{pert}

In this appendix we summarize the details of a perturbative computation that determines the general 2-point function in 
the $\NN=2$ chiral ring of $SU(N)$ $\NN=2$ SCQCD theory up to 3 loops.  Since the OPE coefficients are completely 
fixed in our conventions, our computation also gives results for the perturbative form of the general 3-point functions in the 
$\NN=2$ chiral ring. As explained in the main text, the 
$\NN=2$ chiral primaries of interest are general multi-trace operators of the form
\beq
\label{pertaa}
\phi_{\{n_s\} } = \NN_{\{ n_s \} } \prod_{s=1}^{N-1} \left( \Tr \left[ \varphi^{s+1} \right]\right)^{n_s}
\eeq
where $\NN_{\{ n_s \} }$ are constant normalization factors that will be fixed shortly, and $\varphi$ is the adjoint complex
scalar field in the $\NN=2$ vector multiplet. 

By convention, we consider the trace in the fundamental representation of the 
$SU(N)$ gauge group and normalize the Lie algebra generators $T_a$ ($a=1,2,\ldots,N^2-1$) so that
\beq
\label{pertab}
\Tr \left[ T_a T_b \right] = \delta_{ab}
~.
\eeq
The fully antisymmetric symbol $f_{abc}$, and the fully symmetric symbol $d_{abc}$ are defined as usual
\beq
\label{pertac}
f_{abc} = -i \, \Tr \left[ [T_a,T_b] T_c \right]~, ~~
d_{abc} = \Tr \left[ \{ T_a,T_b \} T_c \right]
~.
\eeq

Then, expressing the adjoint complex scalar field $\varphi$ as
\beq
\label{pertad}
\varphi = \varphi^a T_a
\eeq
we can recast the generic chiral primary \eqref{pertaa} of scaling dimension $\Delta$ into the form
\beq
\label{pertae}
\phi_{\{ n_s \}} =\NN_{\{n_s\}}\, \CC_{\{n_s \}; a_1 \cdots a_\Delta} \, \varphi^{a_1} \cdots \varphi^{a_\Delta}~, ~~
\Delta = \sum_{s = 1}^{N-1} (s+1) \, n_s
~,
\eeq
where
\bea
\label{pertaf}
\CC_{\{n_s \}; a_1 \cdots a_\Delta} =&&\big( \Tr \left[ T_{a_1} T_{a_2} \right] \cdots \Tr \left[ T_{a_{2 n_1-1}} T_{a_{2n_1}} \right] \big)
\\
&&\big( \Tr \left[ T_{a_{2n_1+1}} T_{a_{2n_1+2}} T_{a_{2n_1+3}} \right] \cdots 
\Tr \left [T_{a_{2n_1+3n_2-2}} T_{a_{2n_1+3n_2-1}} T_{a_{2n_1+3n_2}} \right]  \big) \cdots
\nonumber
\eea
is the obvious product of traces of Lie algebra generators.

Following the conventions of Ref.\ \cite{Baggio:2014sna} at tree level the 2-point function of the adjoint scalar components
$\varphi^a$ is 
\beq
\label{pertag}
\big \langle \varphi^a (x) \, \overline{\varphi}^b (0) \big \rangle 
= \delta^{ab} \frac{1}{\pi \,{\rm Im} \tau } \frac{1}{|x|^2}
~.
\eeq
We fix the constant normalization factors $\NN_{\{ n_s \}}$ of the operators $\phi_{ \{ n_s \} }$ so that (after the standard
Wick contractions) these operators have tree-level 2-point functions
\beq
\label{pertah}
\big \langle \phi_{\{ n_s \} }(x) \overline{\phi}_{\{ n_s \}} (0) \big \rangle 
= \frac{1}{(4\, {\rm Im}\tau )^\Delta} ~ \CC_{\{n_s\}; a_1 \cdots a_\Delta} 
\sum_{\sigma \in \SS_\Delta} \CC_{\{n_s\}; a_{\sigma(1)} \cdots a_{\sigma(\Delta)}} 
\frac{1}{|x|^{2\Delta}}
~,
\eeq
where $\SS_\Delta$ is the permutation group of $\Delta$ elements.
This choice is consistent with the normalization that leads to the \tts equations \eqref{introa4}; in particular, it is consistent 
with the OPE
\beq
\label{pertak}
\phi_{\{ n_s \}} \cdot \phi_{\{ m_s \}} \sim \phi_{\{n_s+m_s \}}
~.
\eeq

It is convenient to compute perturbative corrections to correlation functions in the $\NN=2$ SCQCD theory using 
supergraph methods in $\NN=1$ superspace language. 
In fact, the relevant computation of 2-point functions in the $\NN=2$ chiral ring up to order $\OO(\gym^4)$ in the Yang-Mills coupling
$g$ is quite similar to a 3-loop computation of 2-point functions of chiral primary operators in $\NN=4$ SYM theory performed 
previously in \cite{Penati:2000zv}. As expected by the known non-renormalization theorems, and verified explicitly in 
\cite{Penati:2000zv}, the correction in $\NN=4$ SYM theory vanishes. Hence, it is convenient to perform the $\NN=2$ SCQCD
computation by subtracting the corresponding contributions of the analogous computation in $\NN=4$ SYM theory (the same 
approach in this context was employed successfully in the past using standard Feynman diagrams in real space in 
\cite{Andree:2010na,Baggio:2014sna}).

In $\NN=1$ superspace language the 2-point functions of interest take the form
\beq
\label{pertal}
\big \langle \phi_{\{ n_s \}}(z_1) \, \overline{\phi}_{\{\barM n_s \}}(z_2) \big \rangle
= \frac{F( \{ n_s\}, \{ \barM n_s \},\gym^2)}{(x_1-x_2)^{2\Delta}} \delta^{(4)} (\theta_1 - \theta_2)
\eeq
where $z= (x,\theta,\barM\theta)$ are superspace coordinates. We are after the perturbative form of the spacetime-independent
factor $F$
\beq
\label{pertam}
F = F_0 + \gym^2 F_2 + \gym^4 F_4 +\OO(\gym^6)
~.
\eeq

In the $SU(N)$ $\NN=2$ SCQCD theory besides the $\NN=1$ vector superfield $V$ and the adjoint chiral superfield $\varphi$ 
we have $N_f=2N$ fundamental doublets of chiral superfields $Q_i, \tilde Q_i$. Following closely the 
superspace conventions of Ref.\ \cite{Penati:2000zv} (with the obvious additional features of $\NN=2$ SCQCD compared to 
$\NN=4$ SYM) we have four types of super-propagators
\bea
\label{pertan}
V~{\rm propagator} ~~&:& ~~ \includegraphics[trim = 0cm 14cm 0cm 14cm,clip,scale=0.25]{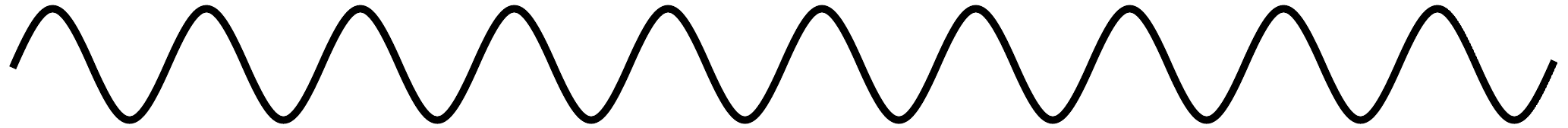} 
\\
\varphi ~ {\rm propagator} ~~&:&~~ \includegraphics[trim = 0cm 14cm 0cm 14cm,clip,scale=0.25]{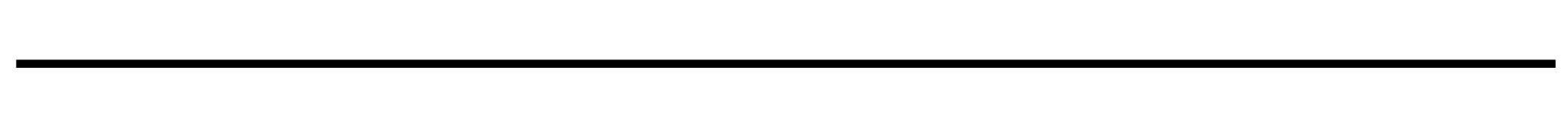} 
\\
Q_i~ {\rm propagator}~~ &:&~~ \includegraphics[trim = 0cm 14cm 0cm 14cm,clip,scale=0.25]{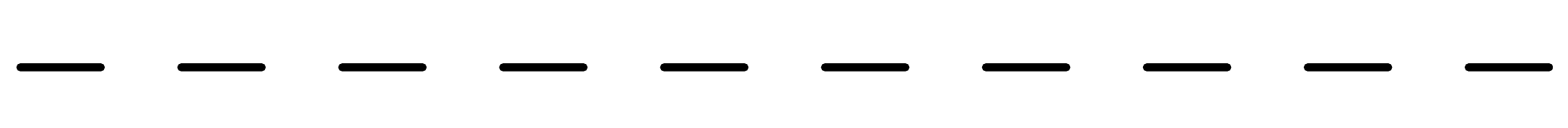} 
\\
\tilde Q_i ~ {\rm propagator}~~ &:& ~~\includegraphics[trim = 0cm 14cm 0cm 14cm,clip,scale=0.25]{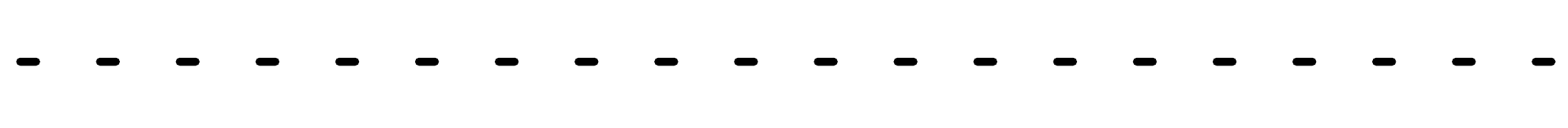}
\eea
There are also eight types of super-vertices 
\bea
\label{pertao}
&& \includegraphics[trim = 0cm 5cm 0cm 5cm,clip,scale=0.15]{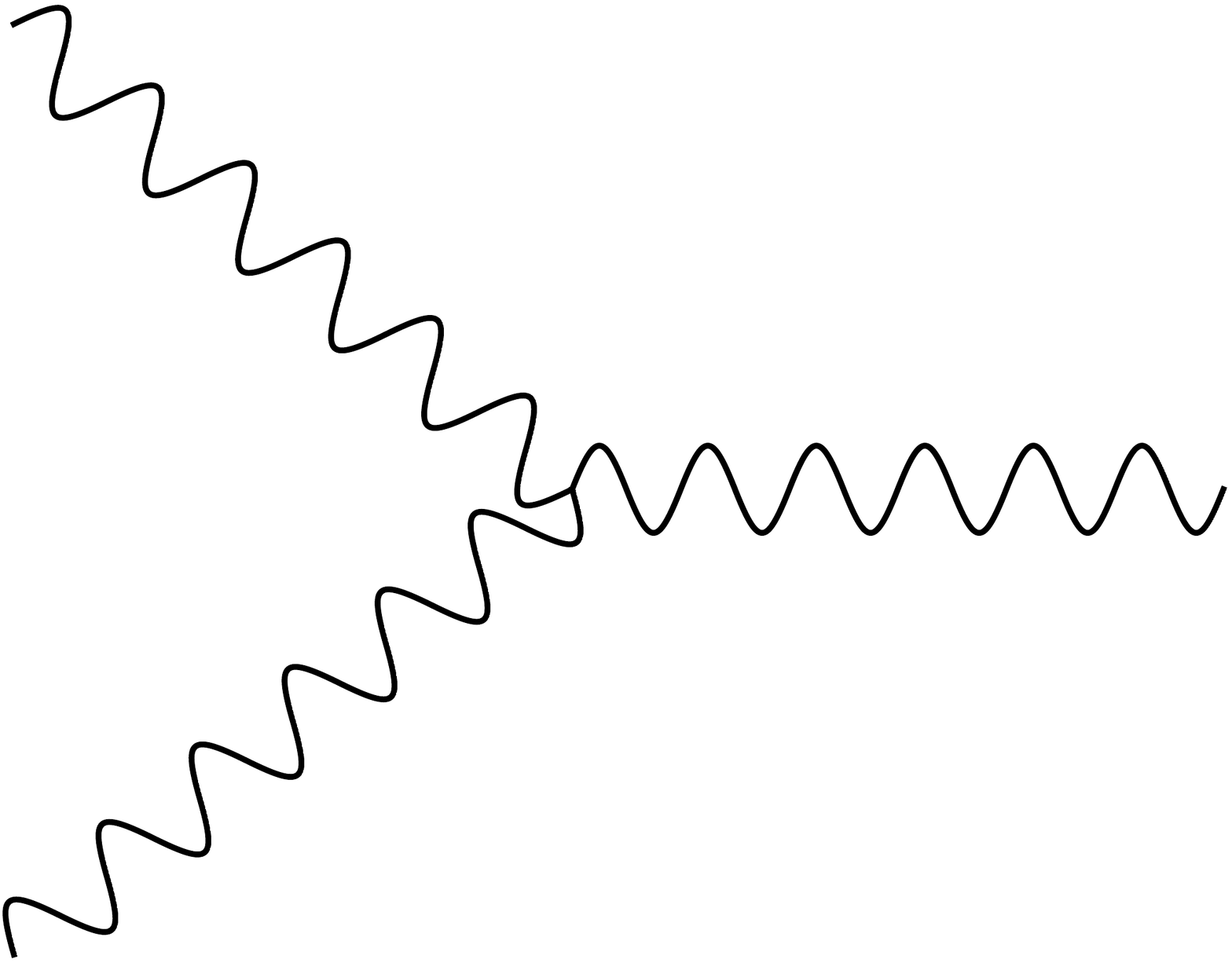} ~~ ~~ 
\includegraphics[trim = 0cm 5cm 0cm 5cm,clip,scale=0.15]{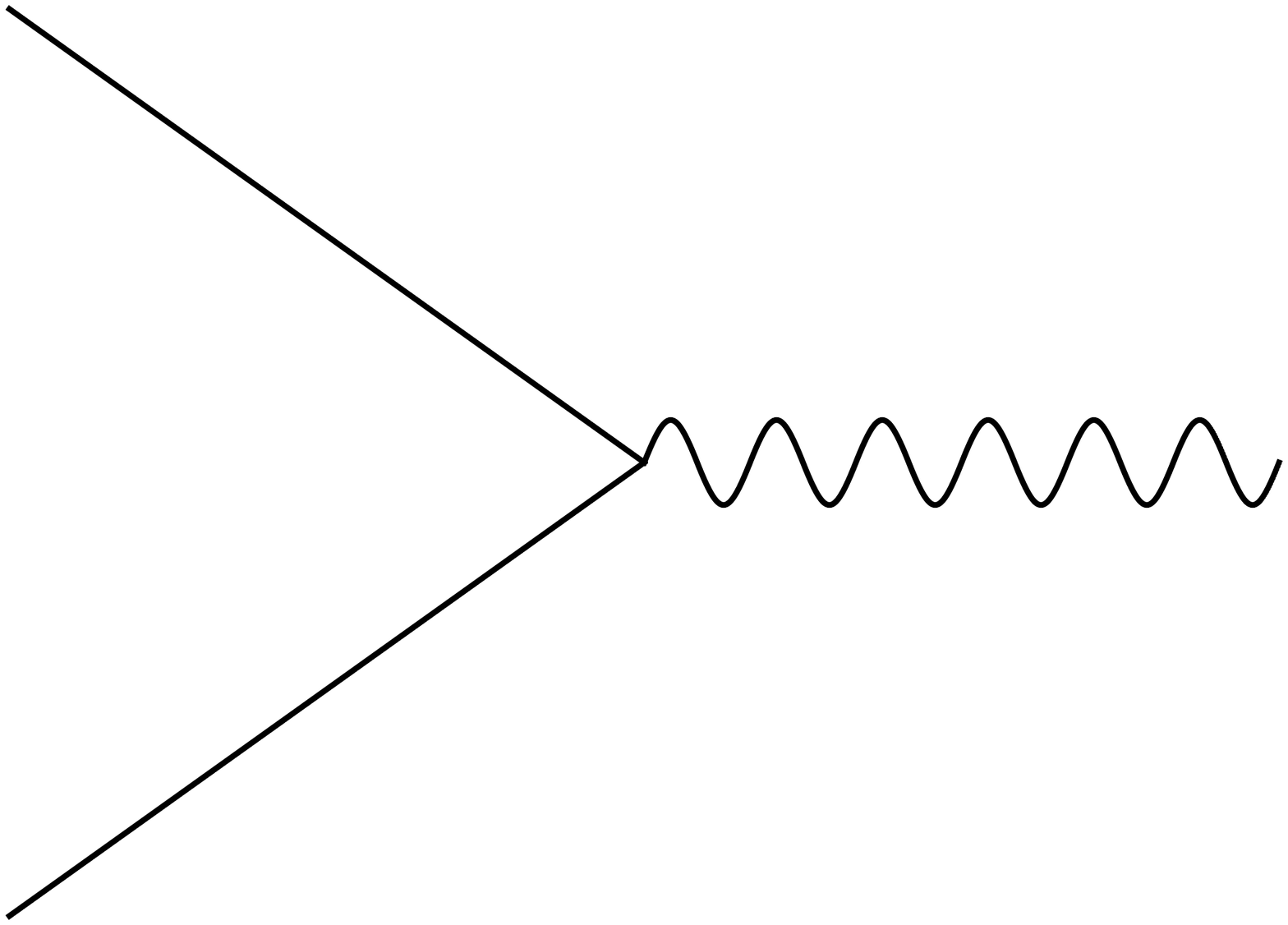} ~~ ~~
\includegraphics[trim = 0cm 5cm 0cm 5cm,clip,scale=0.15]{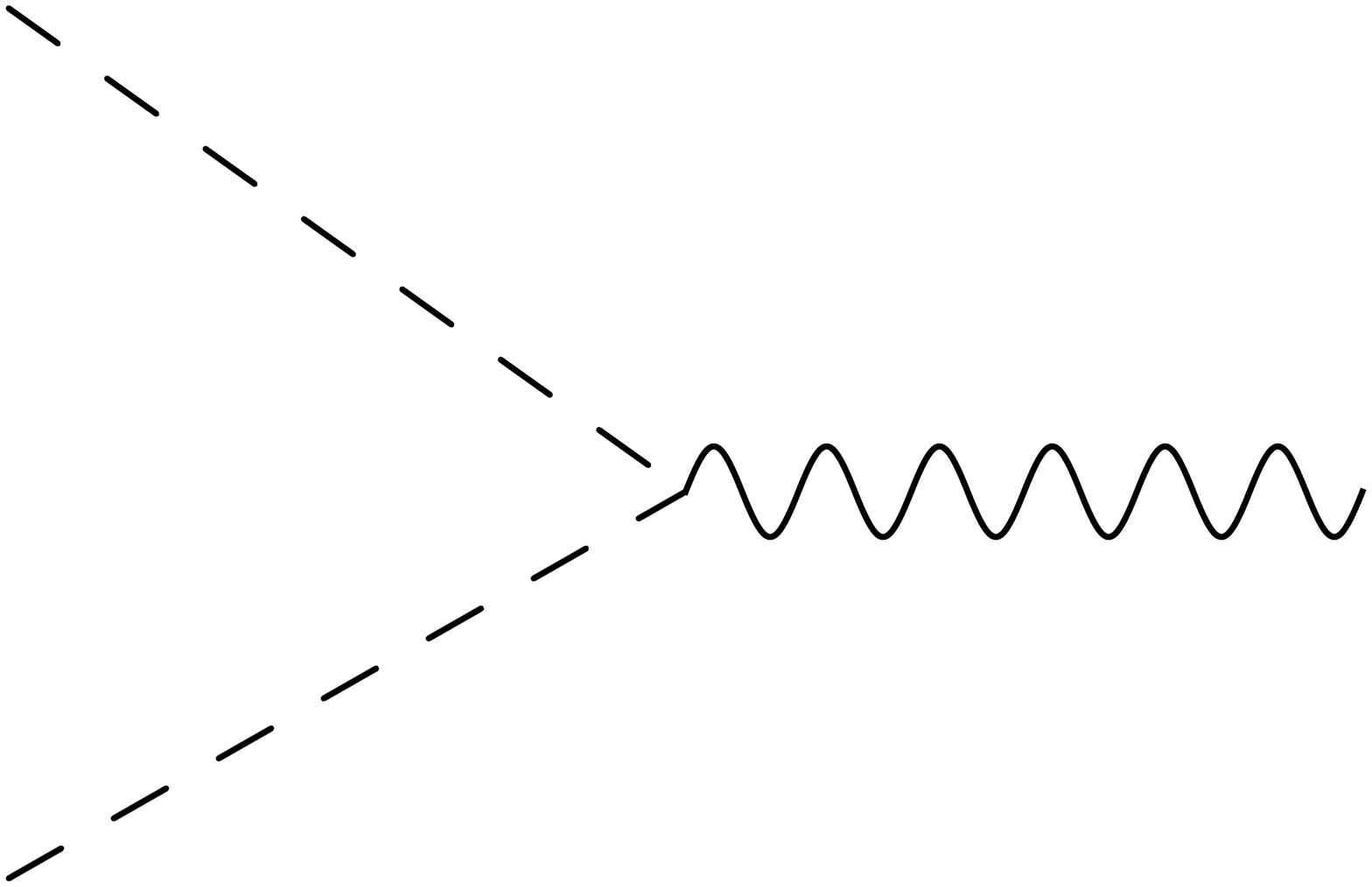}
\nonumber\\
&& \includegraphics[trim = 0cm 5cm 0cm 5cm,clip,scale=0.15]{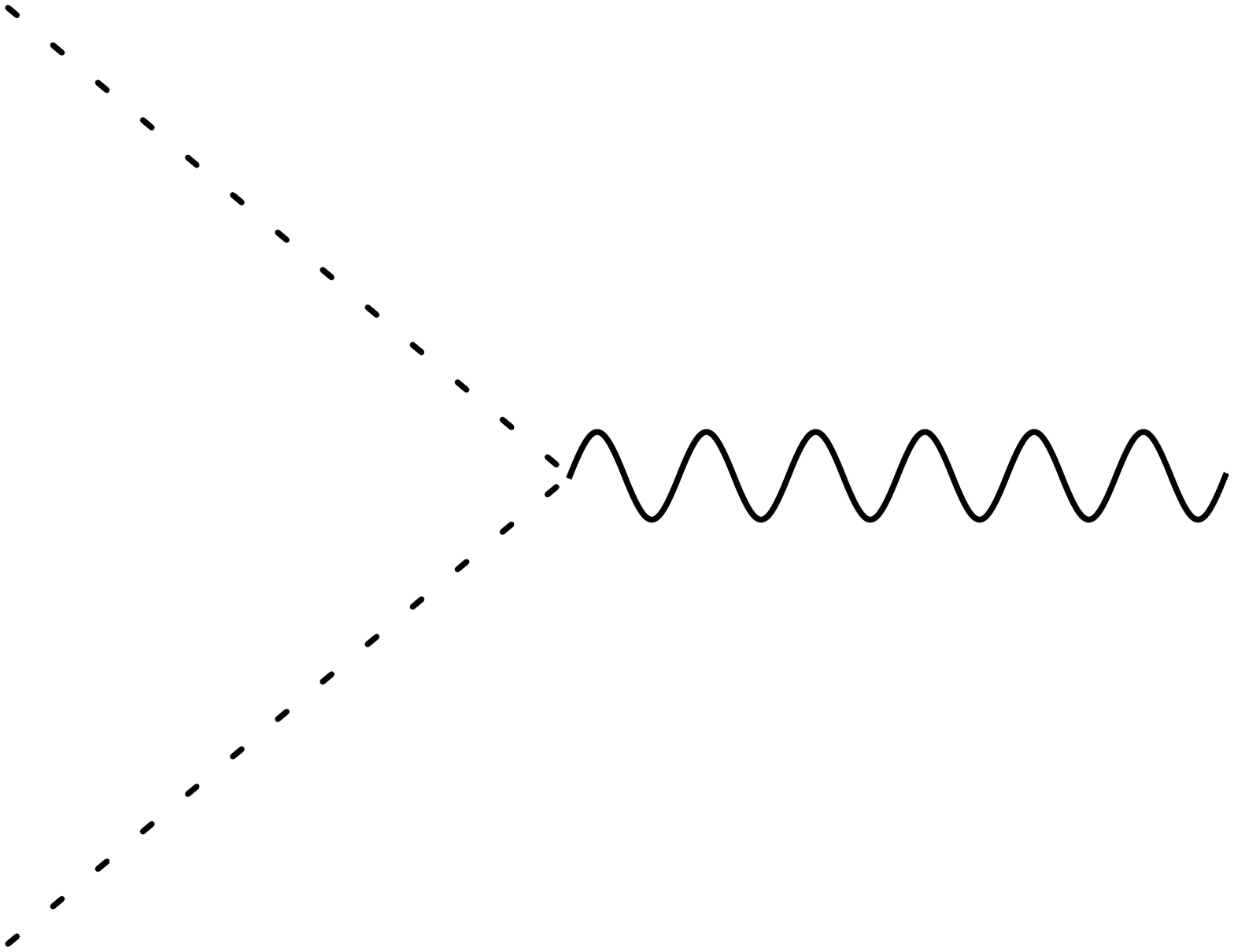} ~~ ~~
\includegraphics[trim = 0cm 7cm 0cm 5cm,clip,scale=0.18]{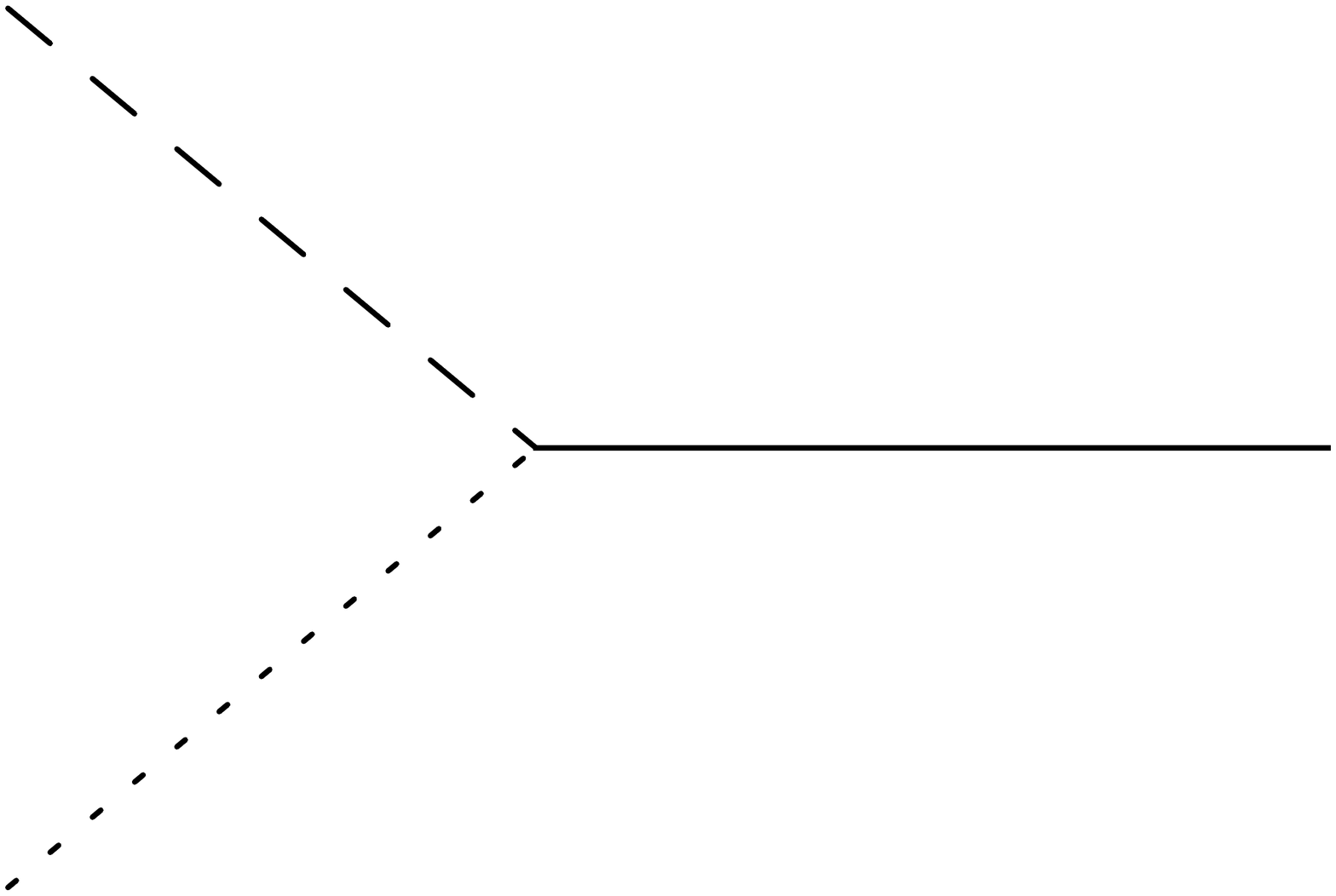} ~~ ~~
\includegraphics[trim = 0cm 3cm 0cm 5cm,clip,scale=0.12]{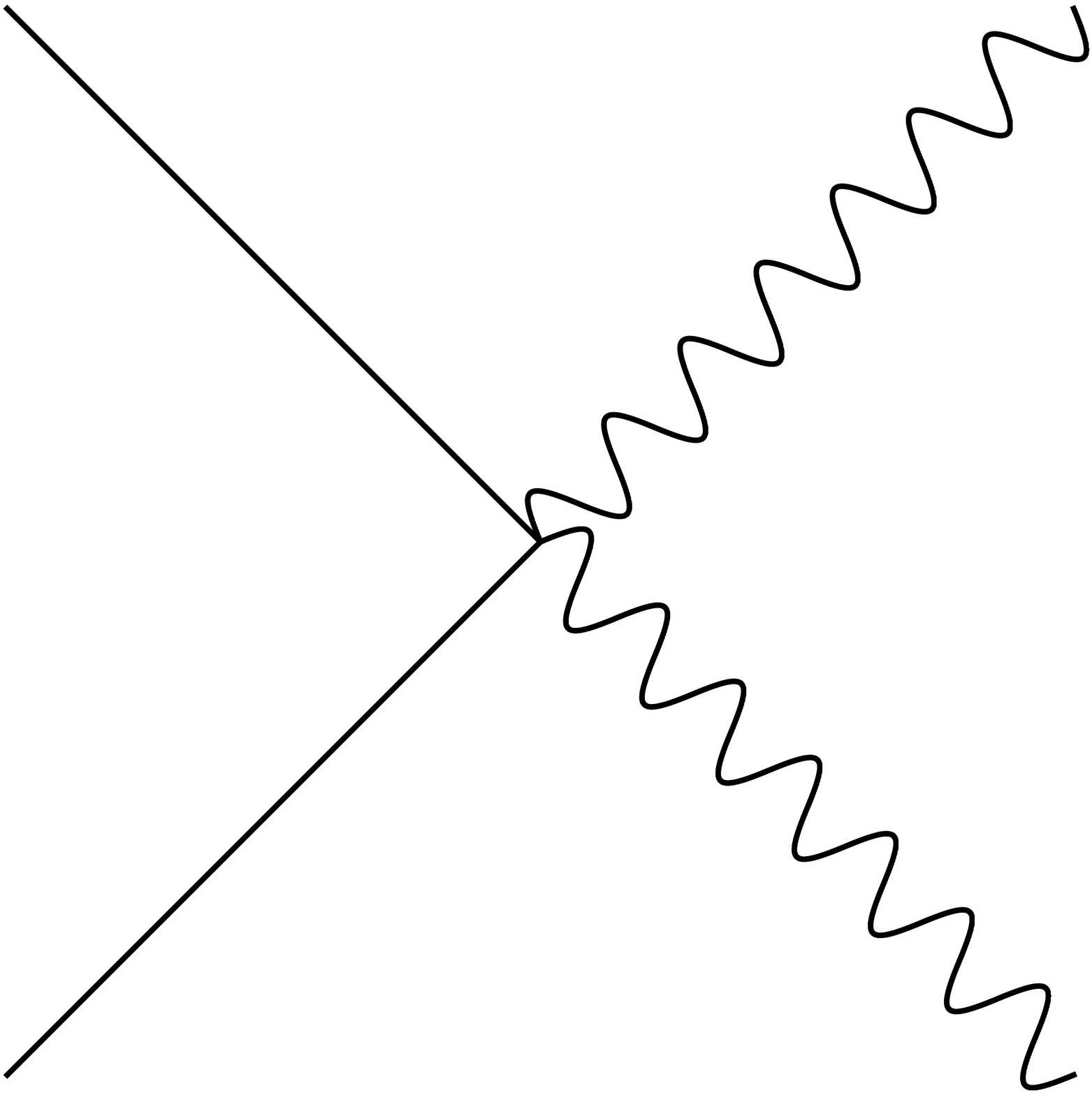}
\nonumber\\
&&~~~~~~~~~~~~ \includegraphics[trim = 0cm 5cm 0cm 5cm,clip,scale=0.15]{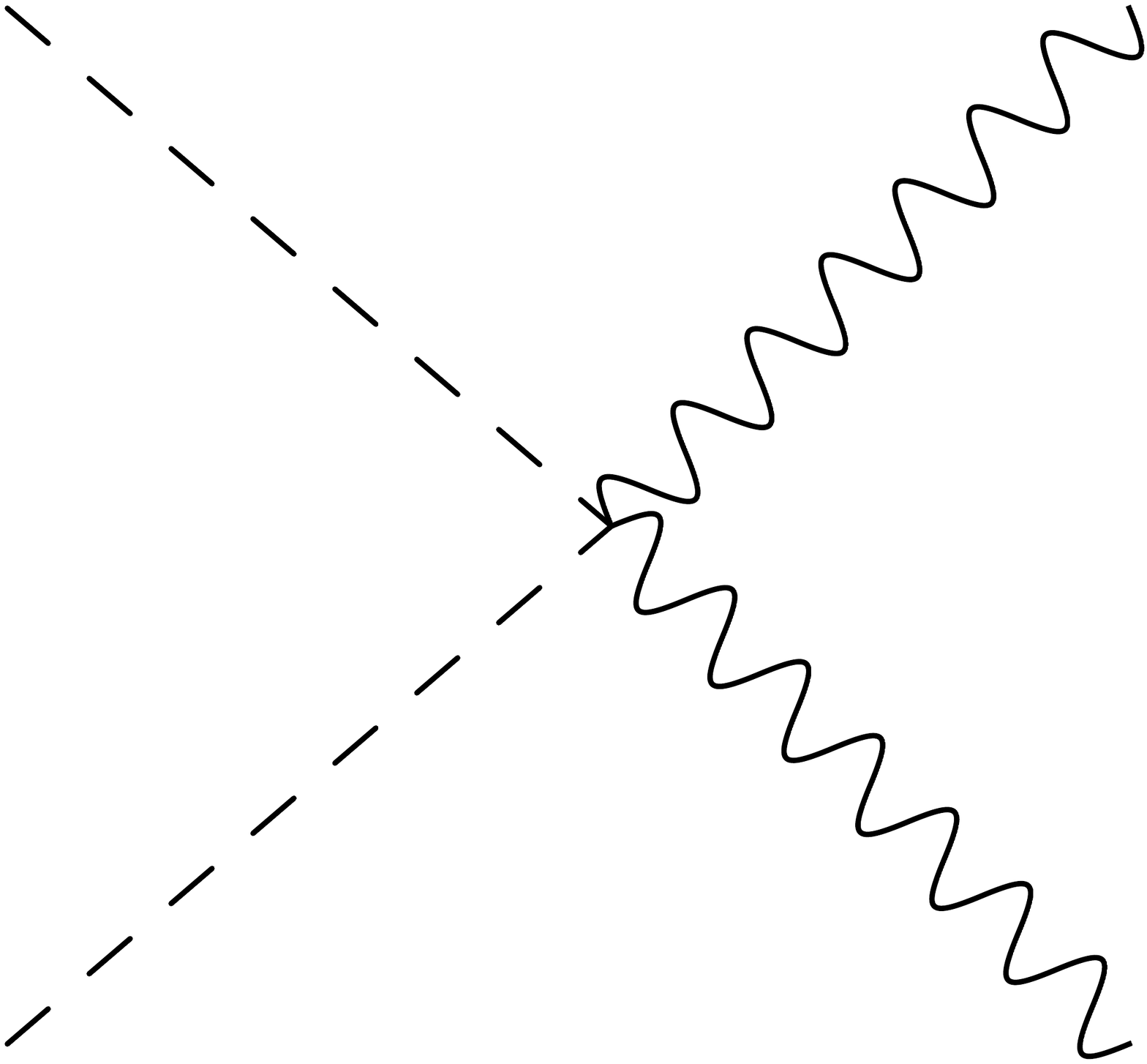} ~~ ~~~~~
\includegraphics[trim = 0cm 5cm 0cm 5cm,clip,scale=0.15]{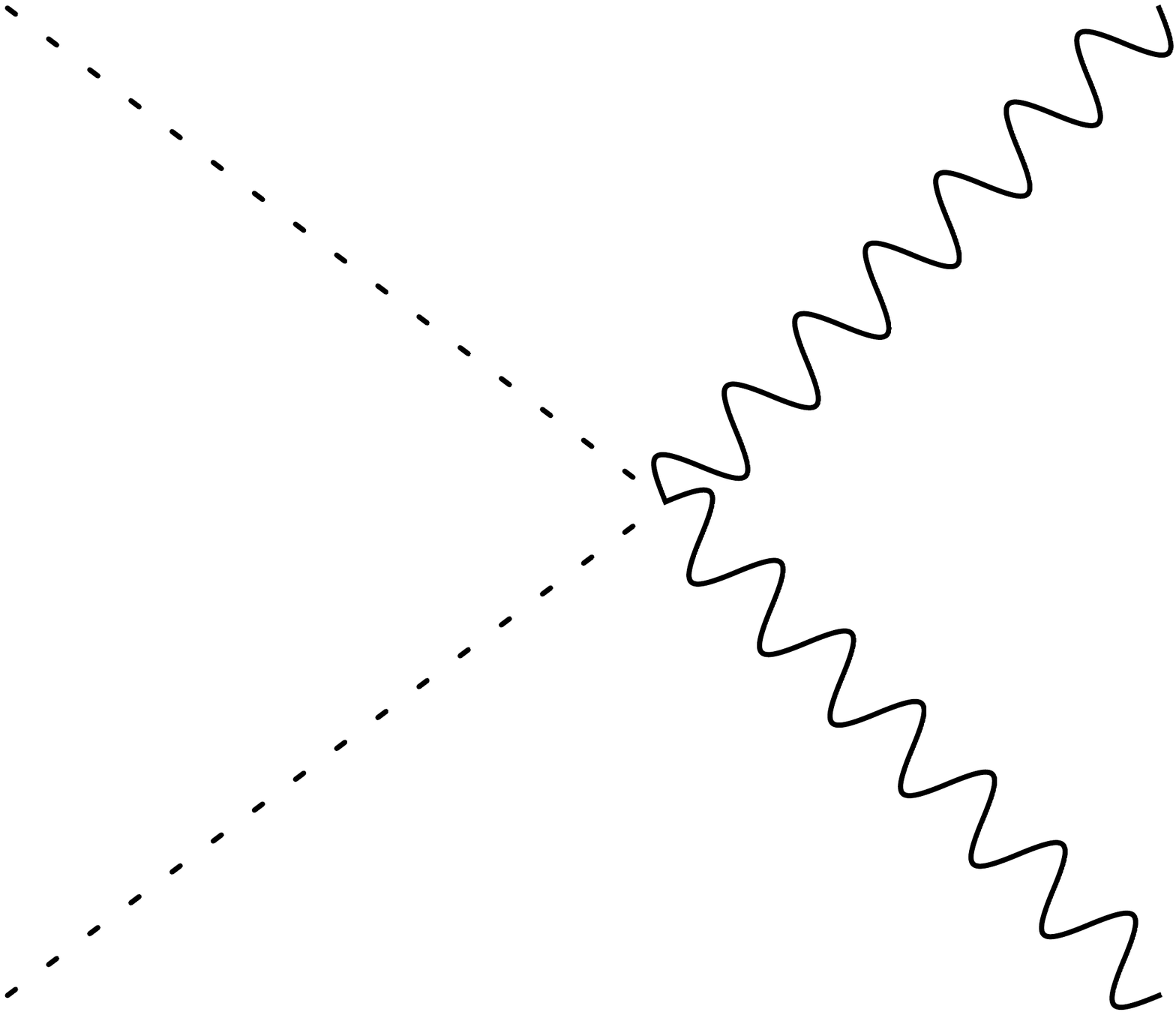}
\eea

We will now sketch how different contributions to the function $F$ arise up to 3 loops in perturbation theory 
highlighting the differences from the $\NN=4$ SYM case (a detailed exposition of several needed facts can be found in 
Ref.\ \cite{Penati:2000zv}). 

At tree level the non-color factor is evaluated from the super-Feynman diagram 
\beq
\label{pertap}
\includegraphics[trim = 5cm 8cm 0cm 15cm,clip,scale=0.6]{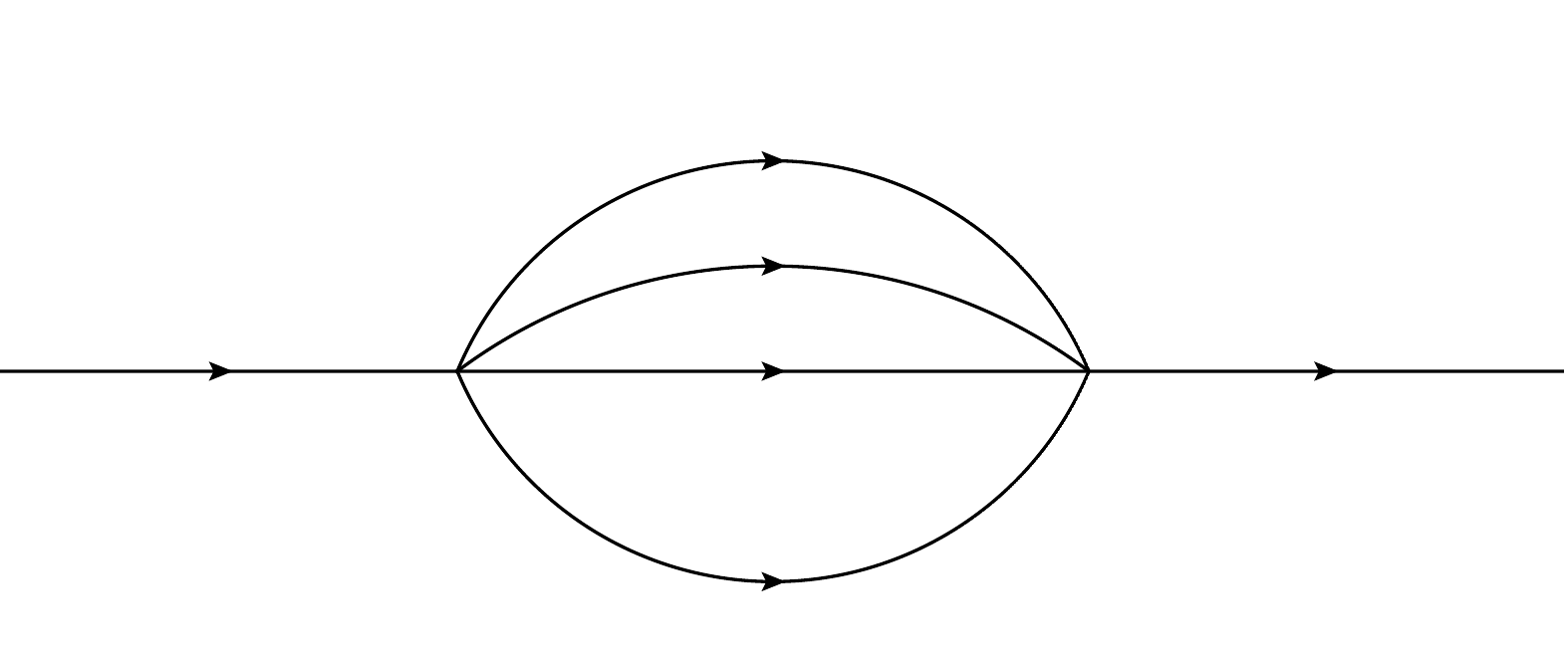}
\eeq
as in $\NN=4$ SYM theory. In our conventions the result is 
\beq
\label{pertapa}
F_0 = \frac{1}{(4\, {\rm Im}\tau )^\Delta} ~ \CC_{\{n_s\}; a_1 \cdots a_\Delta} 
\sum_{\sigma \in \SS_\Delta} \CC_{\{\barM n_s\}; a_{\sigma(1)} \cdots a_{\sigma(\Delta)}} 
\eeq
in agreement with equation \eqref{pertah}.

At the next order, $\OO(\gym^2)$, the only potential contribution comes from diagrams of the form
\beq
\label{pertaq}
\includegraphics[trim = 0cm 8cm 0cm 8cm,clip,scale=0.3]{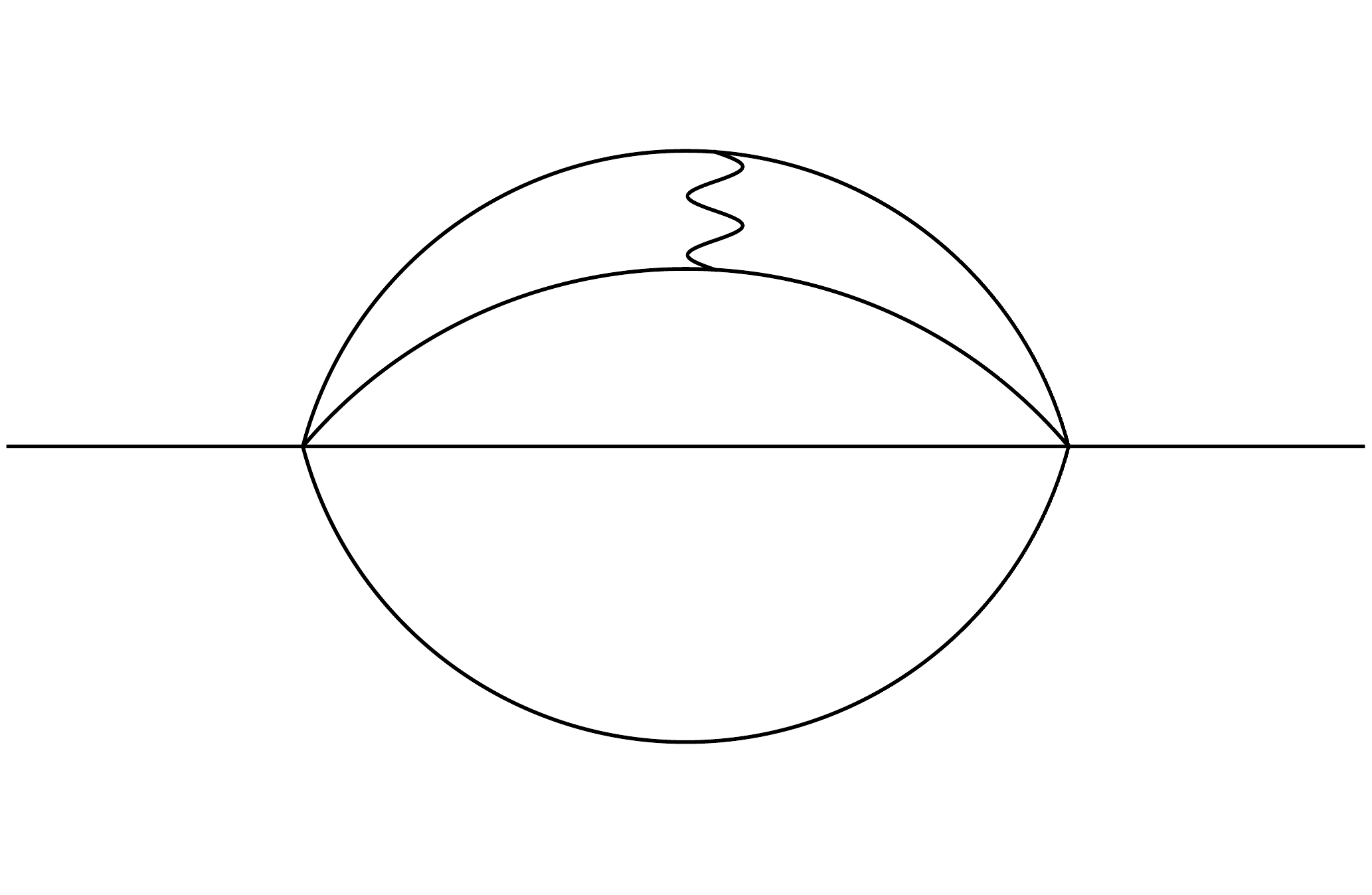}
\eeq
However, as explained in \cite{Penati:2000zv} none of these diagrams give a requisite $1/\varepsilon$ pole in dimensional 
regularization, and as a result, there is no contribution to $F_2$. Namely,
\beq
\label{pertar}
F_2 = 0
~.
\eeq

The first non-vanishing correction arises at order $\OO(\gym^4)$. Besides the diagrams that are common with $\NN=4$ SYM theory
(and will not be listed here) the contributing diagrams to the non-color factor in $\NN=2$ SCQCD theory at this order are
\begin{itemize}
\item[(1)]
\beq
\label{pertasa}
\includegraphics[trim = 0cm 8cm 0cm 8cm,clip,scale=0.3]{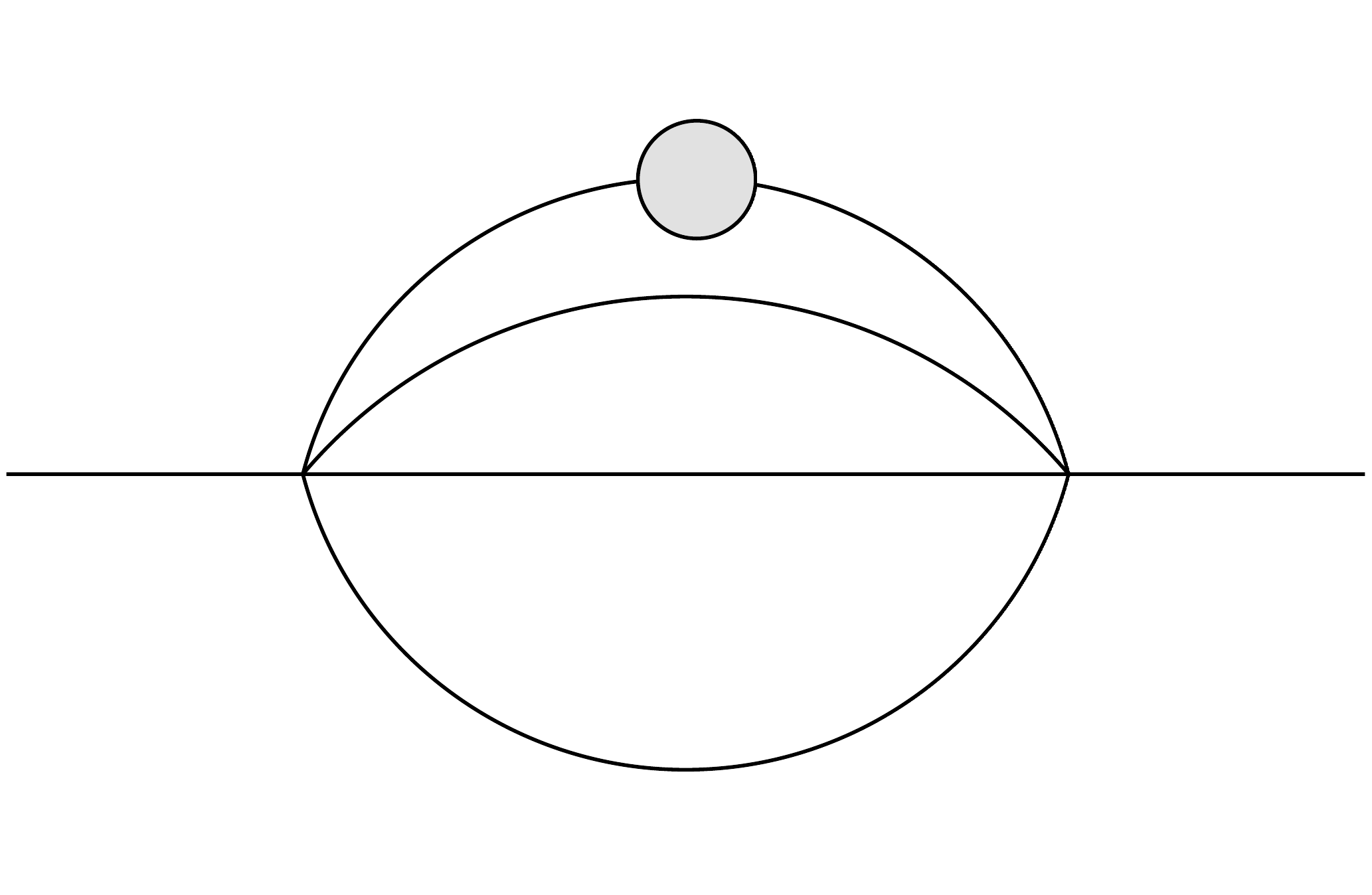}
\eeq
that involves the 2-loop corrected $\varphi$-propagator,
\item[(2)]
\beq
\label{pertasb}
\includegraphics[trim = 6cm 7cm 0cm 14cm,clip,scale=0.45]{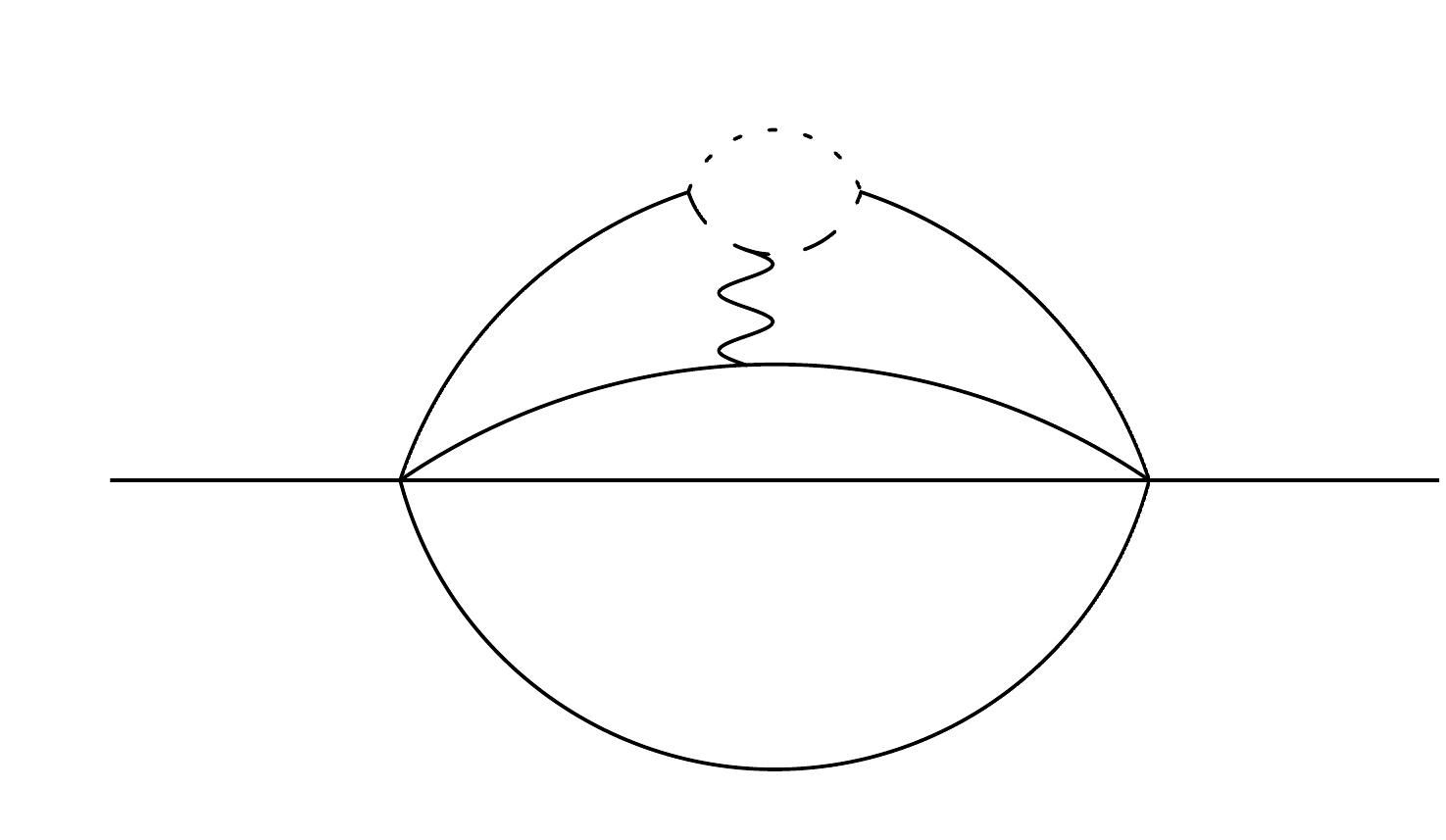}
\includegraphics[trim = 6cm 7cm 0cm 14cm,clip,scale=0.45]{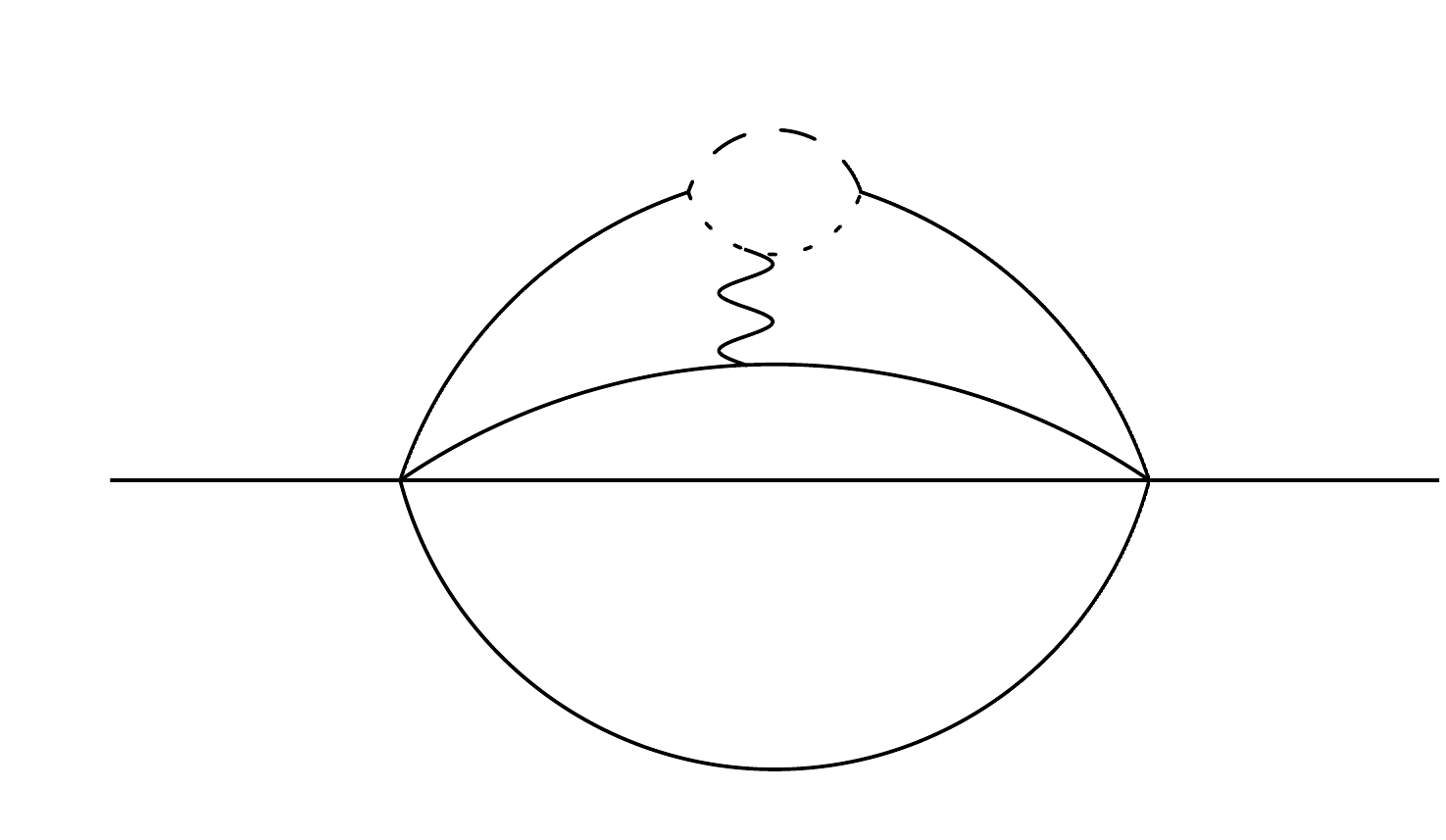}
\eeq
that correct the effective $\overline{\varphi}\varphi V$ vertex,
\item[(3)]
\beq
\label{pertasc}
\includegraphics[trim = 6cm 7cm 0cm 15cm,clip,scale=0.5]{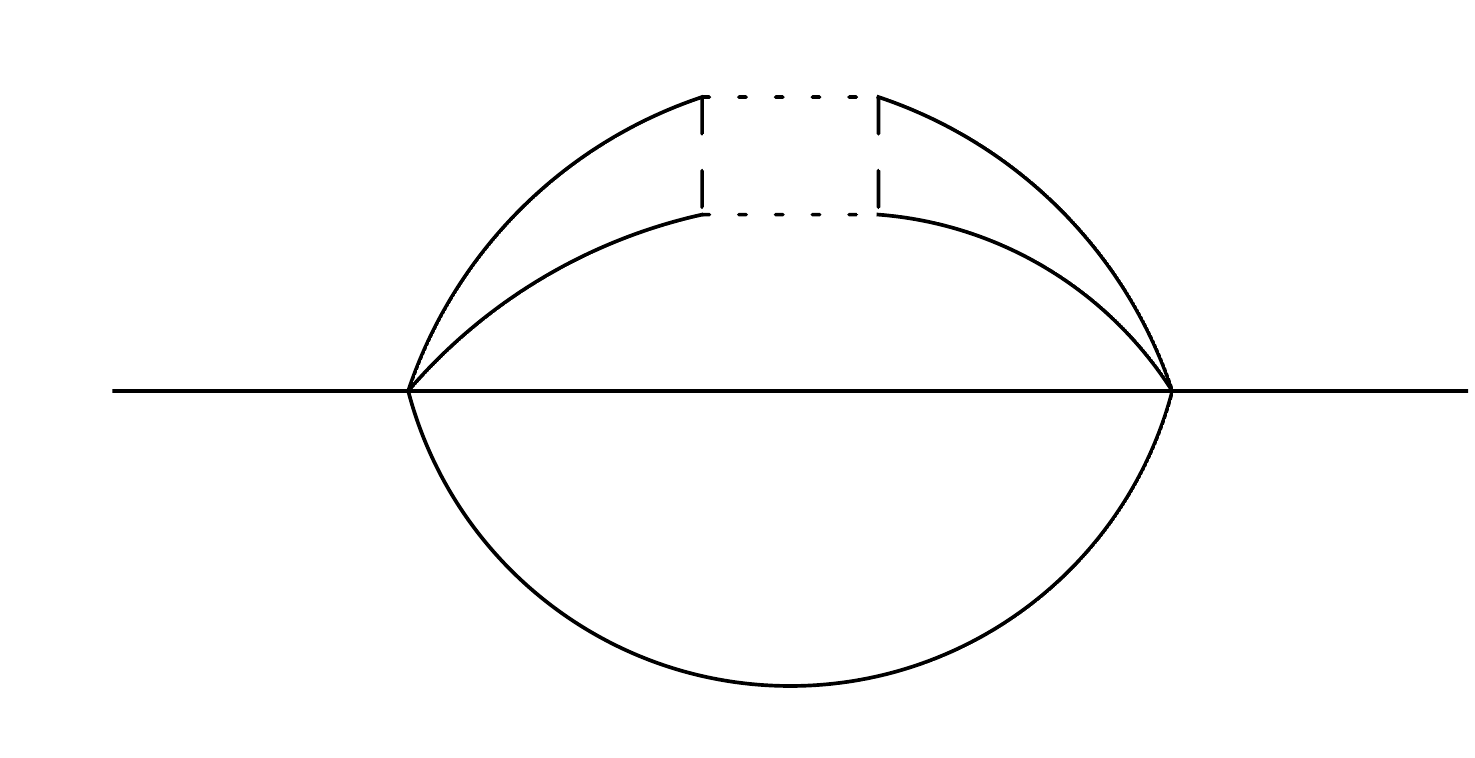}
\includegraphics[trim = 6cm 7cm 0cm 15cm,clip,scale=0.5]{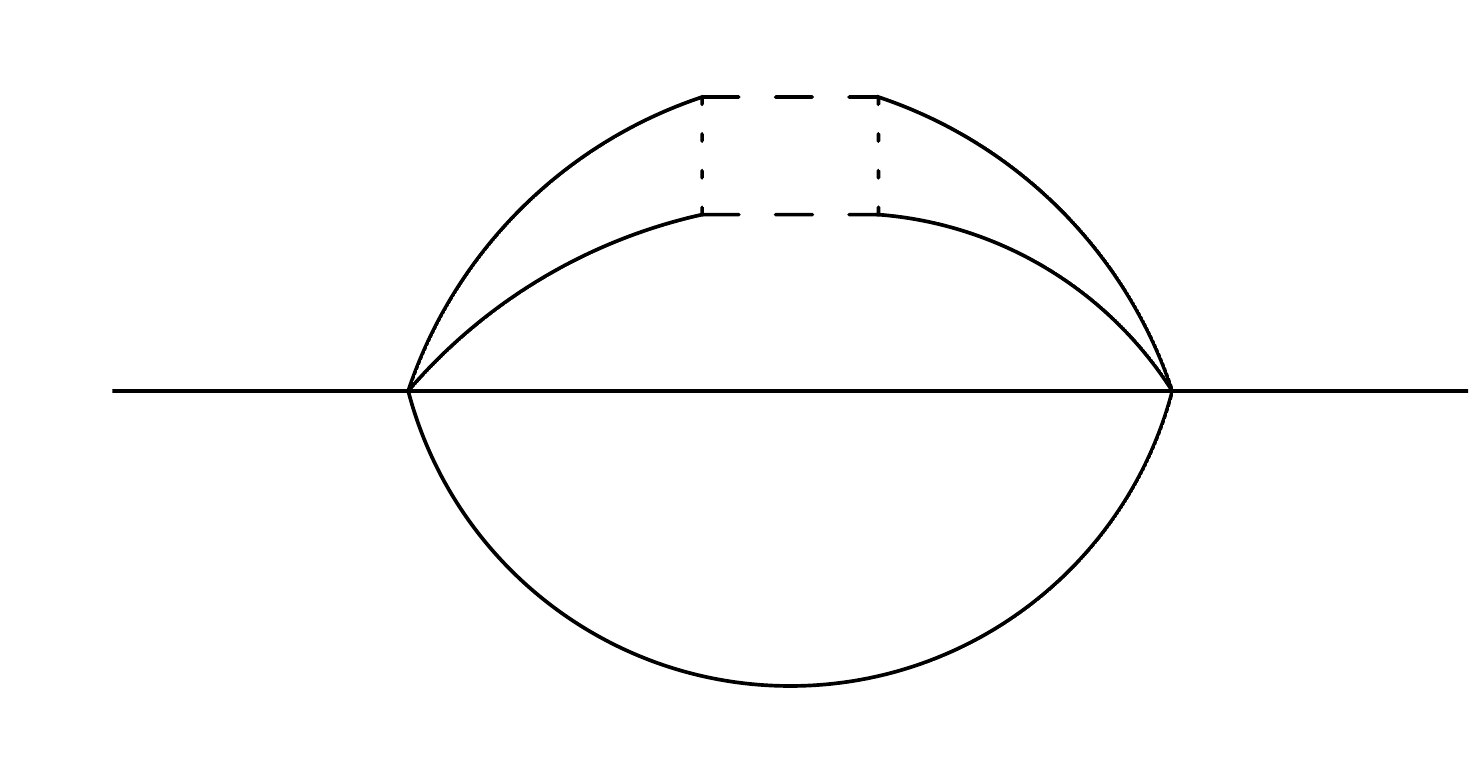}
\eeq
\item[(4)]
\beq
\label{pertasd}
\includegraphics[trim = 6cm 7cm 0cm 15cm,clip,scale=0.5]{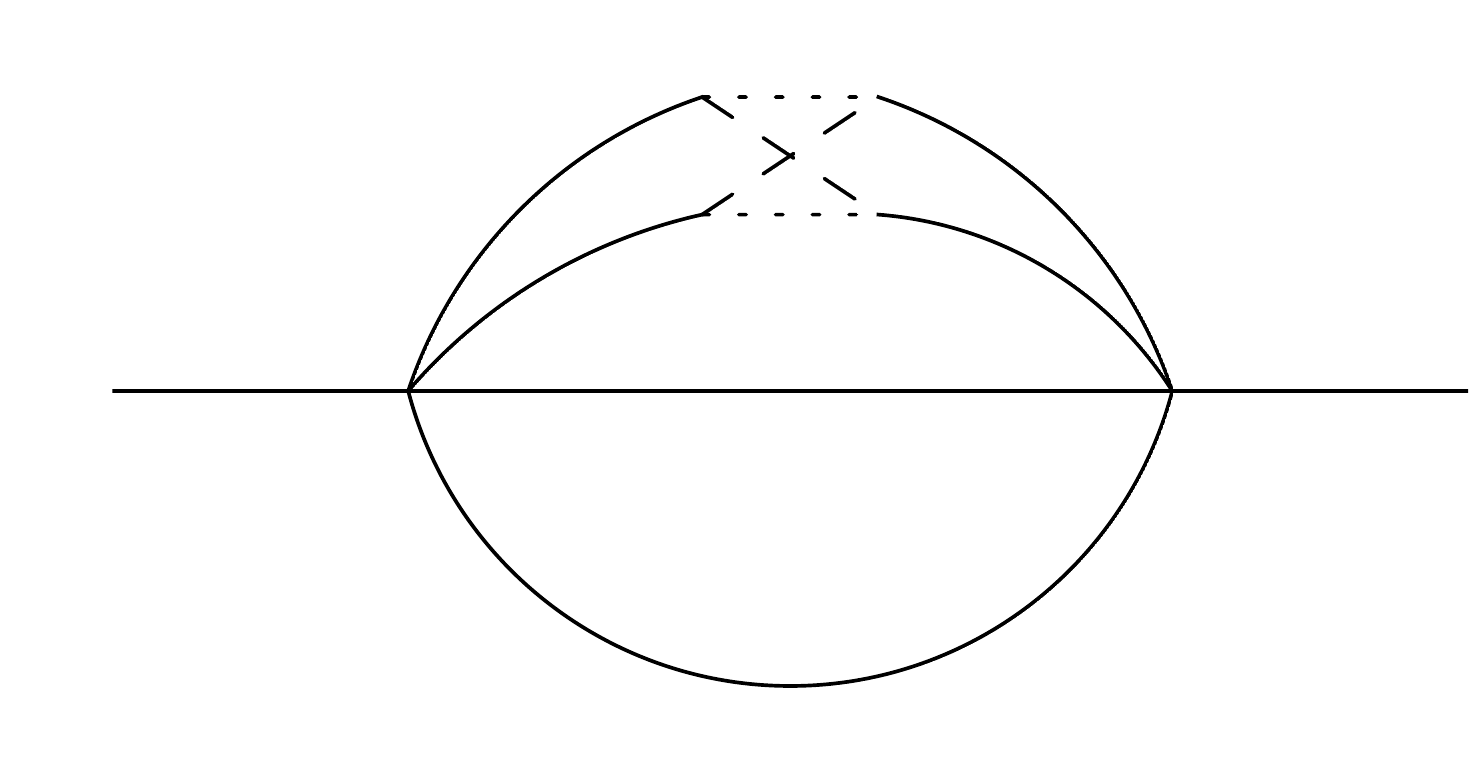}
\includegraphics[trim = 6cm 7cm 0cm 15cm,clip,scale=0.5]{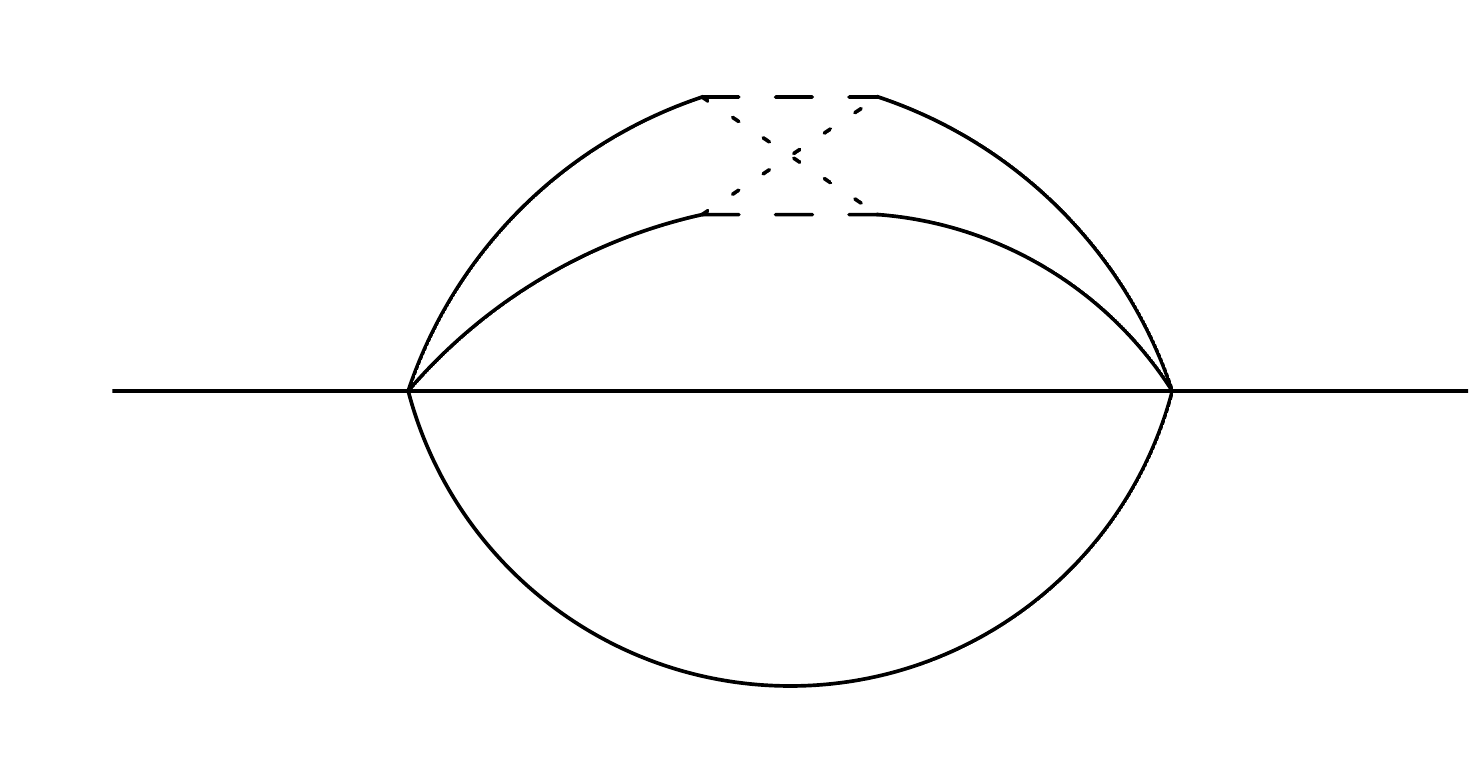}
\eeq
\end{itemize}

The diagrams $(3)$, like the diagrams $3f, 3g,3h$ in \cite{Penati:2000zv} do not contribute to the 
2-point functions. Hence, summing the contributions of the diagrams $(1), (2), (4)$ we get
\beq
\label{pertat}
F_4= F_4^{(1)} + F_4^{(2)} + F_4^{(4)}
~.
\eeq

The difference $(\NN=2)-(\NN=4)$ between the $\NN=2$ and $\NN=4$ results for the 2-loop corrected propagator
is \cite{Andree:2010na}\footnote{Note that compared to equation (18) of \cite{Andree:2010na} in our Lie algebra 
conventions the RHS of the equation is $2(N^2+1)$ versus their $\frac{N^2+1}{2}$.}
\beq
\label{pertba}
12 \, \,\zeta(3) \, \gym^4 \, (N^2+1) \, \frac{1}{(p^2)^{2\varepsilon}}
~.
\eeq
Then, performing the combinatorics and the D-algebra of the full diagram precisely as in \cite{Penati:2000zv} we obtain 
\beq
\label{pertbb}
F_4^{(1)} = - \left( \frac{1}{4\pi} \right)^4 \left( \frac{1}{4\, {\rm Im}\tau} \right)^\Delta 
12 \Delta (N^2+1) \,\zeta(3)~
\CC_{\{n_s \};a_1 \cdots a_\Delta}
\sum_{\sigma \in \SS_\Delta} \CC_{\{\barM n_s\};a_{\sigma(1)} \cdots a_{\sigma(\Delta)} }
~.
\eeq

Similarly, we can easily deduce the $(\NN=2)-(\NN=4)$ difference for the effective $\overline{\varphi}\varphi V$ vertex 
\bea
\label{pertbc}
&&\includegraphics[trim = 0cm 12cm 0cm 10cm,clip,scale=0.2]{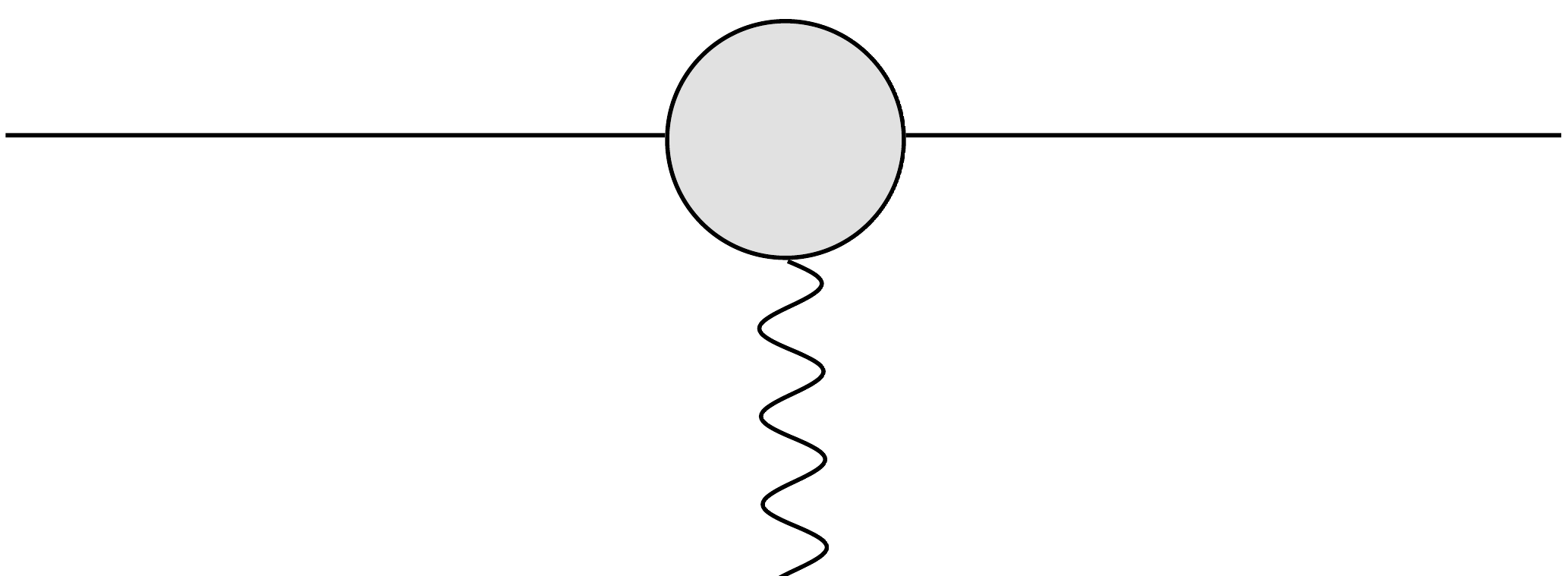}
^{\NN=2} ~~{- \atop } ~~
\includegraphics[trim = 0cm 12cm 0cm 10cm,clip,scale=0.2]{vertex}
^{\NN=4} ~~{= \atop }
\\
&&\frac{N g^3}{4} d_{abc} \overline{\varphi}^a (q,\theta) \varphi^a (-p,\theta)
\Big( 4 D^\alpha \barM D^2 D_\alpha + (p+q)^{\alpha \dot \alpha} \left[ D_\alpha, \barM D_{\dot \alpha} \right]\Big)
V^c (p-q,\theta) \int \frac{ d^n k}{k^2 (k-p)^2 (k-q)^2}
~.\nonumber
\eea
Doing the full D-algebra as in \cite{Penati:2000zv} we finally obtain
\bea
\label{pertbd}
F_4^{(2)} = \left( \frac{1}{4\pi} \right)^4 \left( \frac{1}{4\, {\rm Im}\tau} \right)^\Delta 
12 N \,\zeta(3) ~
\CC_{\{n_s\};a_1\cdots a_\Delta} \sum_\sigma \sum_{j\neq \ell} 
\CC_{\{\barM n_s \};a_{\sigma(1)} \cdots b_j \cdots b_\ell \cdots a_{\sigma(\Delta)} } ~i f_{a_{\sigma(j)} m b_j } 
d_{a_{\sigma(\ell)} m b_\ell}
~.
\nonumber\\
\eea

For the final term $F_4^{(4)}$ we compute only the contribution of the diagrams $(4)$ and subtracting
the contribution of the corresponding $\NN=4$ diagrams with the adjoint superfields running in the loop we find 
\beq
\label{pertbe}
F_4^{(4)} =  \left( \frac{1}{4\pi} \right)^4 \left( \frac{1}{4\, {\rm Im}\tau} \right)^\Delta 
12\,\zeta(3) ~
\CC_{\{n_s\};a_1\cdots a_\Delta} \sum_\sigma \sum_{j\neq \ell} 
\CC_{\{\barM n_s\}; a_{\sigma(1)} \cdots b_j \cdots b_\ell \cdots a_{\sigma(\Delta)} } 
\DD_{b_j a_{\sigma(j)}b_\ell a_{\sigma(\ell)}}
\eeq
where
\beq
\label{pertbf}
\DD_{abcd} = \frac{N}{2} \Tr \left[ T_a T_b T_c T_d \right] - \frac{1}{4} f_{a mn}f_{np d} f_{cpq} f_{qm b}
~.
\eeq

We are now in position to collect the final result for the perturbative correction at 3 loops\footnote{In all explicit $SU(3)$ and $SU(4)$ 
examples that we worked out the term proportional to the symmetric symbol $d_{abc}$ was found not to contribute at the end. It is interesting
to examine if this is a generic property.}
\bea
\label{pertbg}
&&F_4 = \left( \frac{1}{4\pi} \right)^4 \left( \frac{1}{4{\rm Im}\tau} \right)^\Delta
12\,\zeta(3) ~\CC_{\{n_s\};a_1\cdots a_\Delta} 
\\
&& \sum_\sigma
\left[ - (N^2+1) \Delta\, \CC_{\{\barM n_s\};a_{\sigma(1)} \cdots a_{\sigma(\Delta)}} 
+  \sum_{j\neq \ell} \barM \CC_{a_{\sigma(1)} \cdots b_j \cdots b_\ell \cdots a_{\sigma(k)}} 
\, \left( i N f_{a_{\sigma(j)}m b_j} d_{a_{\sigma(\ell)} m b_\ell} + \DD_{b_j a_{\sigma(j)} b_\ell a_{\sigma(\ell)}}\right)
\right]
~.\nonumber
\eea

As a check of these results we have verified that the above formula for $F=F_0+\gym^4 F_4$ reproduces correctly 
the Zamolodchikov metric \cite{Gerchkovitz:2014gta}
\beq
\label{pertbi}
g_2 \equiv \left(\frac{\pi}{4}\right)^2 \big \langle \Tr[\varphi^2](1) \Tr[\overline{\varphi}^2](0) \big \rangle
= \partial_\tau \partial_{\barM \tau} \log Z_{S^4}
\eeq
in the case of the gauge groups $SU(2)$, $SU(3)$, and $SU(4)$, when the exact $S^4$ partition function 
(determined by localization \cite{Pestun:2007rz}) is expanded at this order.

\section{Explicit diagonalization of 2-point functions}
\label{explicit}

\paragraph{Diagonalization of 2-point functions.} 
The diagonalization of the matrix of 2-point function coefficients $g_{K\barM L}$ can be performed in different ways. 
Gram-Schmidt diagonalization is a standard prescription where one picks a first vector $\phi_{K_1}$, then combines
it with a second vector $\phi_{K_2}$ to find a linear combination orthogonal to $\phi_{K_1}$,
then combines $\phi_{K_1}$ and $\phi_{K_2}$ with a third vector $\phi_{K_3}$ to find a linear combination 
orthogonal to the previous two orthogonal vectors and so on and so forth. The choice of the order of the vectors $\phi_{K_1},\ldots$ in this
prescription translates to different linear transformations between the original and the orthogonal bases. 

In what follows, we adopt a slight variant of the Gram-Schmidt diagonalization procedure that reproduces the results of section \ref{c2basis}
based on the $C_2$-algebra \eqref{c2algebra}. 
We single out the first vector $\phi_{K_1}$ in the multi-trace basis \eqref{reviewaaa} as a chiral primary operator with the maximum number
of $\phi_2$ factors. Then, we perform a first linear transformation 
\beq
\label{relaaa}
(\phi^{(1)})_K =  (\MM_1)_K^{~~L} \, \phi_L 
\eeq
that leaves $\phi_{K_1}$ unchanged and transforms all the remaining vectors
to set $\langle \phi^{(1)}_L \barM \phi_{K_1}\rangle =0$ for $L \neq K_1$. 
A general matrix $(\MM_1)_K^{~~L}$ with these properties takes the form
\beq
\label{relaa}
\MM_1 = \left( 
   \begin{matrix} 
      1 & 0 & 0 & \cdots \\
      -\sum_{L\neq K_1} \mu_{L_1 \barM L }\frac{g_{L \barM K_1}}{g_{K_1 \barM K_1}}  & \mu_{L_1 \barM L_1} & \mu_{L_1 \barM L_2} & \cdots \\
      -\sum_{L\neq K_1} \mu_{L_2 \barM L } \frac{g_{L \barM K_1}}{g_{K_1 \barM K_1}}  & \mu_{L_2 \barM L_1} & \mu_{L_2 \barM L_2} & \cdots \\
      \cdots & \cdots & \cdots & \cdots \\
   \end{matrix}
   \right)
   ~
\eeq
where the indices $L_i$ refer to chiral primaries other than $\phi_{K_1}$ and the matrix elements $\mu_{L_i \barM L_j}$ are free
for the moment. We will generate non-trivial entries $\mu_{L_i \barM L_j}$ sequentially, hence at this stage we 
adopt a scheme where $\mu_{L_i \barM L_j} = \delta_{L_i \barM L_j}$. Then,
\beq
\label{relab}
\MM_1 = \left( 
   \begin{matrix} 
      1 & 0 & 0 & \cdots \\
      -\frac{g_{L_1 \barM K_1}}{g_{K_1 \barM K_1}}  & 1 & 0 & \cdots \\
      -\frac{g_{L_2 \barM K_1}}{g_{K_1 \barM K_1}}  & 0 & 1 & \cdots \\
      \cdots & \cdots & \cdots & \cdots \\
   \end{matrix}
   \right)
   ~.
\eeq

At the second step we single out a vector $\phi^{(1)}_{K_2}$ (other than $\phi_{K_1}$), with the next largest number of $\phi_2$ factors, 
and repeat the same transformation in the subspace that excludes $\phi_{K_1}$. Accordingly, we perform a second linear transformation
\beq
\label{relac}
(\phi^{(2)})_K =  (\MM_2)_K^{~~L} \, (\phi^{(1)})_L 
\eeq
with 
\beq
\label{relac}
\MM_2 = \left( 
   \begin{matrix} 
      1 & 0 & 0 & 0 & \cdots \\
      0 & 1 & 0 & 0 & \cdots \\
      0 & -\frac{g^{(1)}_{L_1 \barM K_2}}{g^{(1)}_{K_2 \barM K_2}}  & 1 & 0 &  \cdots \\
      0 & -\frac{g^{(1)}_{L_2 \barM K_2}}{g^{(1)}_{K_2 \barM K_2}}  & 0 & 1 & \cdots \\
      \cdots & \cdots & \cdots & \cdots & \cdots \\
   \end{matrix}
   \right)
   ~.
\eeq
By $g^{(1)}_{K\barM L}$ we have denoted the 2-point function coefficients in the transformed basis $\phi^{(1)}_K$,
\beq
\label{relad}
g^{(1)}_{K\barM L} = g_{K\barM L} - \frac{g_{K \barM K_1} g_{K_1 \barM L}}{g_{K_1 \barM K_1}} 
\eeq
for $K, L \neq K_1$.

We continue in this fashion until the full diagonalization of the matrix $g_{K\barM L}$. The complete transformation matrix is
\beq
\label{relae}
\MM = \MM_{D_R-1} \ldots \MM_2 \MM_1
\eeq
where $D_R$ is the degeneracy of the chiral primary fields with $U(1)_R$ charge $R$. The chiral primaries in the new basis are
\beq
\label{relaea}
\hat \phi_K = \MM_K^{~~L} \phi_L
\eeq
and the matrix of 2-point function coefficients is diagonal
\beq
\label{relaeb}
\langle \hat \phi_K \barM {\hat \phi}_L \rangle = \hat g_{K\barM L} = \hat g_{K\barM K} \delta_{\barM K\barM L}
~.
\eeq

We encounter the same freedom in this process that we encountered also in section \ref{c2basis}. When two operators 
have the same number of $\phi_2$ factors it is unclear which order we should proceed in.

The no-mixing conjecture of section \ref{finitecoupling} postulates that the new section remains holomorphic, i.e.\ the 
linear transformation matrix $\MM$ is a holomorphic function of the moduli
\beq
\label{relaf}
\partial_{\barM \tau} \MM =0
~.
\eeq
This is equivalent to the conditions
\beq
\label{relag}
\partial_{\barM \tau} \MM_i = 0~, ~~ i = 1, 2, \ldots, D_R-1
~,
\eeq 
which translate to 
\beq
\label{relai}
\partial_{\barM \tau} \left( \frac{g^{(i-1)}_{L \barM K_i}}{g^{(i-1)}_{K_i \barM K_i}} \right) = 0~, ~~ i =1,2,\ldots, D_R-1
~.
\eeq
By definition $g^{(0)}_{K\barM L} = g_{K\barM L}$.

As expected by consistency, these relations are invariant under a holomorphic rescaling of the chiral primary fields. 
Moreover, they imply that by suitable holomorphic rescalings
it is possible to adopt a more specific set of normalization conventions where all the 2-point function coefficients $g_{K\barM L}$ are real.
This is the real $\phi_K$ basis that was aluded to in the main text and was explicit in the perturbative computations. 
In this basis the complex conjugate of 
the relations \eqref{relai} implies that the ratio of 2-point functions is also $\tau$-independent. As a result, in the real basis 
the ratios $\frac{g^{(i-1)}_{L \barM K_i}}{g^{(i-1)}_{K_i \barM K_i}}$ are coupling constant independent and their value is 
fixed at tree level, namely
\beq
\label{relaj}
\frac{g^{(i-1)}_{L \barM K_i}}{g^{(i-1)}_{K_i \barM K_i}} = \left( \frac{g^{(i-1)}_{L \barM K_i}}{g^{(i-1)}_{K_i \barM K_i}} \right)_{tree-level}
~.
\eeq
This equation is a statement of non-renormalization formulated in a local patch (based on the holomophic gauge) on the superconformal
manifold of the $\NN=2$ SCQCD theory.

\paragraph{Diagonalization of $C_2$.} 
In the construction of section \ref{c2basis} based on the $C_2$-algebra \eqref{c2algebra} the simultaneous diagonalization of the 
OPE coefficient $C_2$ was automatic. In the above language this property can be formulated as follows. 
In the original basis \eqref{reviewaaa} the normalization conventions guarantee $C_{2 K}^L = \delta^L_{K+2}$. 
After the linear transformation \eqref{relae} we obtain the new OPE coefficients
\beq
\label{relba}
\hat C_{2 K}^L = (\MM_{(\Delta)})_K^{~~S}\, C_{2 S}^P \, (\MM_{(\Delta+2)}^{-1})_P^{~~L} 
= (\MM_{(\Delta)})_K^{~~S} (\MM_{(\Delta+2)}^{-1})_{S+2}^{~~~~~L}
~,
\eeq
where we used the fact that the chiral primary $\phi_2 \propto \Tr[\phi^2]$ is the single scaling dimension 2 operator and does
not transform. Also, we used the notation $\MM_{(\Delta)}$ to denote the tranformation matrix at scaling dimension $\Delta$.
Notice that transformation matrices at two different scaling dimensions appear on the RHS of equation \eqref{relba}. 
It is obvious that the dimensionality of the transformation matrices remains the same or increases as the scaling dimension 
increases, i.e.\ $D_{2\Delta} \leq D_{2(\Delta+2)}$.

Let us phrase the precise conditions that guarantee that the transformed OPE coefficients $\hat C_{2 K}^L$ remain proportional to $\delta^L_{K+2}$.
Part of our prescription above is to organize the rows and columns of the transformation matrix $\MM_{(\Delta+2)}$ so 
that its $i$-th row and column (for $i\leq D_{2\Delta}$) refers to the chiral primary of the $i$-th row and column of $\MM_\Delta$
after the OPE with $\phi_2$. It is then straightforward to verify that
\beq
\label{relbb}
\hat C_{2K}^L = \delta^L_{K+2} ~~\Leftrightarrow ~~ 
\frac{g^{(i-1)}_{L \barM K_i}}{g^{(i-1)}_{K_i \barM K_i}} = \frac{g^{(i-1)}_{L+2~ \barM K_i+\barM 2}}{g^{(i-1)}_{K_i+2~ \barM K_i+\barM 2}} 
~, ~~ i =1,2,\ldots, D_R
~.
\eeq

\paragraph{The above relations from the viewpoint of the \tts equations before the tranformation.} Before we end this appendix, 
we would like to present a slighlty different description of the above relations from the point of the view of the \tts equations 
in the original basis \eqref{reviewaa}. 
Returning to the \tts equations \eqref{introa4} in the multi-trace basis \eqref{reviewaaa} we single out the first chiral primary 
$\phi_{K_1}$ (that takes part in the above diagonalization procedure, see eq.\ \eqref{relaaa}), and 
consider the component of the equations with $K=K_1$ and $L\neq K_1$
\beq
\label{decoupleab}
\partial_{\barM \tau} \left( g^{\barM M L} \partial_\tau g_{K_1 \barM M} \right)
= g_{K_1+2,\barM R + \barM 2}\, g^{\barM R L} 
- g_{K_1 \barM R} \, g^{\barM R - \barM 2, L-2}
~.
\eeq
It follows easily from the previous discussion that the non-renormalization equations \eqref{relaj} set the LHS (connection part) of this equation to zero.
The RHS vanishes as a consequence of equations \eqref{relbb}. Indeed, using these equations
\beq
\label{decoupleac}
\frac{g_{K_1+2, \barM M+\barM 2}}{g_{K_1+2, \barM K_1 +\barM 2}} = \frac{g_{K_1 \barM M}}{g_{K_1 \barM K_1}}
= \frac{g_{K_1-2, \barM M-\barM 2}}{g_{K_1-2, \barM K_1 -\barM 2}} 
\eeq
and we can recast the RHS in the form
\beq
\label{decouplead}
{\rm RHS} = \frac{g_{K_1 + 2, \barM K_1+\barM 2} }{g_{K_1 \barM K_1}} g_{K_1 \barM R} \, g^{\barM R L}
- \frac{g_{K_1, \barM K_1} }{g_{K_1-2, \barM K_1-\barM 2}} g_{K_1-2, \barM R-\barM 2}\, g^{\barM R-\barM 2, L-2}
=0
\eeq
since $g_{K_1 \barM R}g^{\barM R L}=0$, $g_{K_1-2, \barM R-\barM 2}g^{\barM R-2, L-2}=0$.

To proceed with the remaining \tts equations, one can perform the transformation \eqref{relaaa}, decouple the chiral 
primary $\phi_{K_1}$, repeat the same argument for $\phi^{(1)}_{K_2}$ with the remaining \tts equations, and so on and so forth.

\bibliographystyle{utphys}
\bibliography{ttstarpaper}
\end{document}